\documentclass[seceq]{ptptex}
\usepackage{wrapft}

\usepackage{graphicx}
\usepackage{eepic}
\usepackage{epic}

\newcommand{\beq}{\begin{equation}}
\newcommand{\eeq}{\end{equation}}
\def\bea{\begin{eqnarray}}
\def\eea{\end{eqnarray}}
\def\nn{\nonumber}

\title{
Systematic Investigation of Possibilities \\
for New Physics Effects in $b\rightarrow s$ Penguin Processes%
}

\author{
C.S. \textsc{Kim}$^{1}$\footnote{cskim@yonsei.ac.kr},
Y.J. \textsc{Kwon}$^1$\footnote{yjkwon63@yonsei.ac.kr},
Jake \textsc{Lee}$^2$\footnote{jilee@cskim.yonsei.ac.kr}
and T. \textsc{Yoshikawa}$^3$\footnote{tadashi@eken.phys.nagoya-u.ac.jp}%
}

\inst{
$^1$Department of Physics, Yonsei
              University, Seoul 120-749, Korea\\
$^2$Center for Quantum Spacetime, Sogang University,
        Seoul 121-742, Korea\\
$^3$Department of Physics, Nagoya University,
                            Nagoya 464-8602, Japan
}

\abst{%
\noindent Although recent experimental results in $b \to s$
penguin process seem to be roughly consistent with the standard
model predictions, there may be still large possibilities of
new physics  hiding in this processes.
Therefore, here we  investigate
systematically the potential new physics effects that may appear in
time-dependent CP asymmetries of $B \rightarrow \phi K^0$,
$B\rightarrow \eta^\prime K^0$ and $B\rightarrow K^0 \pi^0$ decay
modes, by classifying the cases for the values of the mixing-induced
indirect CP asymmetries, $S_{\phi K^0},~ S_{\eta^\prime K^0},~
S_{K^0 \pi^0}$ which are compared to $S_{J/\psi K^0}$. We
also show that several $B_s$ decay modes may help to resolve the
ambiguities in such an analysis. Through combining analysis with
the time-dependent CP asymmetries of $B_s$ decay modes such as
$B_s \rightarrow \phi \eta^\prime $, $B_s\rightarrow \eta^\prime
\pi^0$ and $B_s\rightarrow K^0 \overline{K}^0$, we can determine
where the new CP phases precisely come from.
}

\begin{document}

\maketitle

\section{Introduction}

According to the experimental results from Belle~\cite{Belle} and
BaBar~\cite{BaBar} collaborations which were presented before the
Lepton-Photon Conference 2005 (LP05), there appeared to be a large
discrepancy between the time-dependent CP asymmetries extracted
from $B\rightarrow J/\psi K $ and those from $b \to s$ penguin
processes. Based on the standard model (SM) predictions, where no
significant differences are expected among those processes, this
apparent experimental discrepancy may provide an evidence of new
physics (NP) effects~\cite{newP} in the CP asymmetries beyond what
is understood by the Cabibbo-Kobayashi-Maskawa (CKM) mechanism
\cite{KM}. Hence many scenarios of new physics have been
investigated by many authors~\cite{newP}. With the newest results
\cite{Abe-LP05} reported at LP05, the discrepancy is substantially
reduced, and the CP asymmetries of $B\rightarrow \phi K $ mode
seem to be almost consistent with the SM predictions. However, it
does not mean that the possibility of NP has completely
disappeared at all. According to the results for the other $b\to
s$ penguin processes, there still remain some hints of
discrepancies. Therefore, we investigate more carefully to find
the NP effects hiding in the $B$ decays.

The time-dependent CP asymmetry is defined as follows~\cite{BS}:
\bea \frac{\Gamma({\overline{B}_{phys}^0(t)\rightarrow f_{CP}}) -
      \Gamma({{B}_{phys}^0(t)\rightarrow f_{CP}}) }
     {\Gamma({\overline{B}_{phys}^0(t)\rightarrow f_{CP}}) +
      \Gamma({{B}_{phys}^0(t)\rightarrow f_{CP}}) } &=&
     A_f \cos(\Delta m t) + S_f \sin(\Delta m t ) ,
\eea
where
\bea A_f &=& \frac{|A(\overline{B}^0 \rightarrow f_{CP})|^2 -
              |A({B}^0 \rightarrow f_{CP})|^2 }
             {|A(\overline{B}^0 \rightarrow f_{CP})|^2 +
              |A({B}^0 \rightarrow f_{CP})|^2 } , \\
S_f &=&  \sqrt{ 1 - A_f^2 }  Im \left[ e^{-2i\phi_M}
           \frac{ A({B}^0 \rightarrow f_{CP})^* }
                {|A({B}^0 \rightarrow f_{CP})|}
           \frac{ A(\overline{B}^0 \rightarrow f_{CP}) }
                {|A(\overline{B}^0 \rightarrow f_{CP})|} \right].
\eea
Here $A_f$ indicates the direct CP violation in the decay
$B^0\rightarrow f_{CP}$, and $S_f$ describes a {\em
mixing-induced} indirect CP violation due to the interference
between $B^0-\overline{B}^0$ mixing and its decay process. And
$\phi_M$ represents a weak phase in the $B^0-\overline{B}^0$
mixing. Within the SM the mixing phase $\phi_M$ is just a CKM
phase $\phi_1$ for $B^0_d$ system, whereas it is almost zero\footnote{
We neglect the tiny weak phase existing in $B_s - \bar{B}_s$
system. In more accurate analysis the effects may need to consider
carefully, but this paper is not still in such situation.}
for $B^0_s$ system. If there is no direct CP asymmetry and only the SM
phase exists in the $B^0-\overline{B}^0$ mixing, then the mixing
induced indirect CP asymmetry, with $A_f = 0$, becomes
\bea \label{eq:Sf}
S_f = \left\{ \begin{array}{cc}
               \sin(2\phi_1 + 2 \phi_D) & \;\mbox{ for } B^0_d \mbox{ system,}\\
               \sin(2 \phi_D) & \;\mbox{ for } B^0_s \mbox{ system,}
               \end{array}\right.
\eea
where $\phi_D $ is a weak phase in the decay amplitude defined by
$2 \phi_D = {\rm Arg}\frac{A({B}^0 \rightarrow f_{CP})} {A(\overline{B}^0
\rightarrow f_{CP})} $.

In Table~\ref{CPdata}, we list recent experimental results
\cite{Abe-LP05, Belle-LP05, HFAG} of the
time-dependent CP asymmetries for several relevant modes.
It appears that for the indirect CP asymmetry, $S_f$,
there are apparent differences between $B\to J/\psi K$ mode and
other $b\to s$ penguin-dominated modes such as $B\to \eta' K^0$,
at a level of two standard deviations.
On the other hand, the direct CP asymmetries are consistent with
zero in all cases.

\begin{wraptable}{l}{9cm}
\label{CPdata}
\begin{center}
\begin{tabular}{|c|c|c|c|}\hline
f  & $A_f$ & $S_f$ & Br($B\rightarrow f$)$\times10^6 $ \\
\hline
$J/\psi K^0 $ & -0.027$\pm$0.028 & $0.685\pm0.032$  & $850 \pm 50$ \\[2mm]
$\phi K^0 $
             & $0.09\pm0.14$
                          & $0.47\pm0.19$
                          & $8.3{}^{+1.2}_{-1.0}$   \\[2mm]
$\eta^\prime K^0 $
             & $0.07 \pm 0.07$
             & $0.50 \pm 0.09$
                          & $68.6\pm4.2$  \\[2mm]
$K^0 \pi^0 $
             & $0.02 \pm 0.13$
                          & 0.31 $\pm$ 0.26
                          & $11.5\pm1.0$   \\
\hline
\end{tabular}
\caption{ The experimental results of
          direct ($A_f$) and indirect ($S_f$) CP asymmetries for each
          mode. Averaged values between Belle and
          BaBar are listed~\cite{Abe-LP05,Belle-LP05,HFAG}. }
\end{center}
\end{wraptable}

Motivated by those experimental results, here we
investigate possible NP effects in $b\rightarrow s $ penguin
processes, in particular, $B\rightarrow \phi K $, $B\rightarrow
\eta^\prime K $ and $B\rightarrow K^0 \pi^0$,
which are relatively
easy to be compared with many measurable processes diagrammatically.
Using the
topological quark diagrammatic decomposition method
\cite{ZCC,GLR1,DDM,BuSi,BFRS,DKL,GRZ}, the amplitudes of the modes can be
expressed\footnote{ More complete and useful expansion for all
charmless $B$ decay modes including
higher order contributions has been shown in, for example,
Ref. \citen{BuSi}. Here, for simplicity, we use the
most simple grammatical decomposition based on  Ref. \citen{GLR1}.}
as follows:
\bea
A(B\rightarrow \phi K ) &=& ( \tilde{P}^s + \tilde{S}^s
                - \frac{1}{3}\tilde{P}_{EW}^s )
                        V_{tb}^* V_{ts}, \\
A(B\rightarrow \eta^\prime K^0 ) &=& \frac{1}{\sqrt{6}}\left[ \left\{
     ( {P} + 2 {S} + \frac{1}{3}{P}_{EW} )
 + 2 ( {P}^s + {S}^s - \frac{1}{3}{P}_{EW}^s )
                       \right\} V_{tb}^* V_{ts} \right. \nn \\
   & & ~~~~~~~~~~~~~~~~~~~~~~~~~~~~~~~~~~~~~~~~~~~~~~~~~~~~
         \left.  + ~C ~~V_{ub}^* V_{us} \right], \\
A(B\rightarrow K^0 \pi^0) &=& \frac{1}{\sqrt{2}} \left[
     ( {P} - {P}_{EW} ) V_{tb}^* V_{ts}
               - C V_{ub}^* V_{us} \right] .
\eea
Here, $P$ and $S$ represent $b \to s(\overline q q)$ color-favored
and $b \to s(\overline q q \rightarrow$ SU(3) singlet meson)
singlet QCD penguin diagrams, respectively. $P^s$ and
$S^s$ are the corresponding $b \to s(\overline s s)$ type
diagrams. $C$ stands for color-suppressed tree diagram and
$P^{(s)}_{EW}$ is for electroweak (EW) penguin.
For the parametrization, we follow the method\cite{GLR1},
which is very convenient, and useful to put a hierarchy assumption
among the magnitude of the diagrams within the SM;
$|P|,|P^s| > |S|,|S^s|,|P_{EW}|,|P_{EW}^s|$ and
$|P~V_{tb}^*V_{ts}| >> |C V_{ub}^* V_{us}|$ and  etc.
If SU(3) flavor symmetry is exact, one can find easier relations
between the parameters for $b \to s(\overline q q)$
and $b \to s(\overline s s)$, for example, $P = P^s, S = S^s$.
To decompose
$B\rightarrow \eta^\prime K_s $ decay, we use a choice\cite{GR-eta}
of the quark components of   $\eta^\prime = (u\bar{u}
+d\bar{d}+2s\bar{s})/\sqrt{6}$, which is corresponding to the
octet-singlet mixing angle $\sim 19.5^\circ $\cite{GK}.
We distinguish $B \rightarrow $
vector plus pseudoscalar meson (VP) decays from the
$B$ decays to two pseudoscalar mesons (PP) by adopting the tilde
in corresponding parameters.
In general, the parameters in $B\rightarrow \phi K $ are not
necessarily the same with those in $B\rightarrow$ PP even if the exact
SU(3) flavor symmetry is assumed.

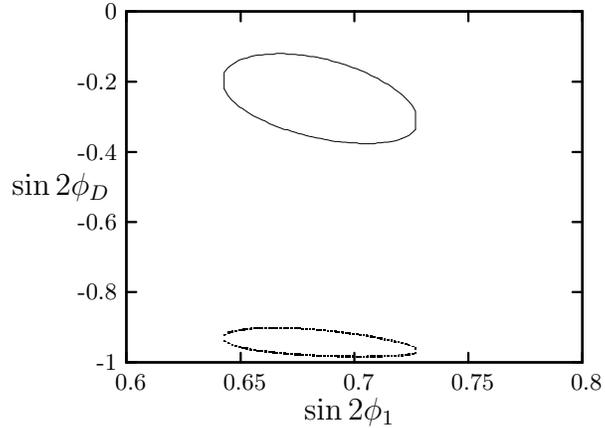
\begin{wrapfigure}{r}{7.5cm}
\begin{center}
\setlength{\unitlength}{0.090450pt}
\begin{picture}(2400,1800)(0,0)
\footnotesize
\thicklines \path(370,249)(411,249)
\thicklines \path(2276,249)(2235,249)
\put(329,249){\makebox(0,0)[r]{-1}}
\thicklines \path(370,543)(411,543)
\thicklines \path(2276,543)(2235,543)
\put(329,543){\makebox(0,0)[r]{-0.8}}
\thicklines \path(370,837)(411,837)
\thicklines \path(2276,837)(2235,837)
\put(329,837){\makebox(0,0)[r]{-0.6}}
\thicklines \path(370,1130)(411,1130)
\thicklines \path(2276,1130)(2235,1130)
\put(329,1130){\makebox(0,0)[r]{-0.4}}
\thicklines \path(370,1424)(411,1424)
\thicklines \path(2276,1424)(2235,1424)
\put(329,1424){\makebox(0,0)[r]{-0.2}}
\thicklines \path(370,1718)(411,1718)
\thicklines \path(2276,1718)(2235,1718)
\put(329,1718){\makebox(0,0)[r]{ 0}}
\thicklines \path(370,249)(370,290)
\thicklines \path(370,1718)(370,1677)
\put(370,166){\makebox(0,0){ 0.6}}
\thicklines \path(847,249)(847,290)
\thicklines \path(847,1718)(847,1677)
\put(847,166){\makebox(0,0){ 0.65}}
\thicklines \path(1323,249)(1323,290)
\thicklines \path(1323,1718)(1323,1677)
\put(1323,166){\makebox(0,0){ 0.7}}
\thicklines \path(1800,249)(1800,290)
\thicklines \path(1800,1718)(1800,1677)
\put(1800,166){\makebox(0,0){ 0.75}}
\thicklines \path(2276,249)(2276,290)
\thicklines \path(2276,1718)(2276,1677)
\put(2276,166){\makebox(0,0){ 0.8}}
\thicklines \path(370,249)(2276,249)(2276,1718)(370,1718)(370,249)
\put(-110,983){\makebox(0,0)[l]{\large $\sin2\phi_D $}}
\put(1323,42){\makebox(0,0){\large $\sin2\phi_1 $  }}
\thinlines \path(778,1394)(778,1394)(794,1366)(810,1347)(825,1335)(841,1319)(857,1310)(873,1298)(889,1289)(904,1279)(920,1273)(936,1264)(951,1258)(967,1249)(982,1243)(998,1237)(1013,1231)(1029,1225)(1044,1222)(1060,1216)(1075,1213)(1090,1207)(1106,1204)(1121,1198)(1136,1195)(1151,1192)(1167,1189)(1182,1186)(1197,1183)(1212,1180)(1227,1177)(1242,1174)(1257,1174)(1272,1171)(1287,1171)(1302,1168)(1317,1168)(1332,1168)(1346,1165)(1361,1165)(1376,1165)(1391,1165)(1405,1165)(1420,1168)(1435,1168)(1449,1171)(1464,1171)(1478,1174)(1493,1177)(1507,1180)(1522,1186)
\thinlines \path(1522,1186)(1536,1192)(1550,1198)(1565,1210)(1579,1225)
\thinlines \dashline[-10]{8}(1579,309)(1579,309)(1565,315)(1550,321)(1536,325)(1522,329)(1507,333)(1493,336)(1478,340)(1464,343)(1449,345)(1435,349)(1420,351)(1405,355)(1391,357)(1376,360)(1361,362)(1346,365)(1332,366)(1317,368)(1302,371)(1287,372)(1272,375)(1257,376)(1242,379)(1227,380)(1212,381)(1197,383)(1182,384)(1167,385)(1151,387)(1136,388)(1121,389)(1106,389)(1090,391)(1075,391)(1060,392)(1044,392)(1029,393)(1013,393)(998,393)(982,393)(967,393)(951,392)(936,392)(920,391)(904,391)(889,389)(873,388)(857,385)(841,384)
\thinlines \dashline[-10]{8}(841,384)(825,380)(810,377)(794,372)(778,363)(778,338)
\thinlines \path(778,1394)(778,1463)(778,1463)(794,1482)(810,1495)(825,1504)(841,1514)(857,1520)(873,1523)(889,1529)(904,1533)(920,1536)(936,1536)(951,1539)(967,1539)(982,1539)(998,1542)(1013,1542)(1029,1542)(1044,1539)(1060,1539)(1075,1539)(1090,1536)(1106,1536)(1121,1533)(1136,1529)(1151,1529)(1167,1526)(1182,1523)(1197,1520)(1212,1517)(1227,1510)(1242,1507)(1257,1504)(1272,1498)(1287,1495)(1302,1488)(1317,1482)(1332,1479)(1346,1472)(1361,1466)(1376,1460)(1391,1450)(1405,1444)(1420,1438)(1435,1428)(1449,1419)(1464,1413)(1478,1403)(1493,1391)(1507,1381)(1522,1369)
\thinlines \path(1522,1369)(1536,1356)(1550,1341)(1565,1325)(1579,1301)(1579,1225)
\thinlines \dashline[-10]{8}(1579,309)(1579,288)(1579,288)(1565,284)(1550,282)(1536,280)(1522,278)(1507,277)(1493,277)(1478,275)(1464,275)(1449,275)(1435,274)(1420,274)(1405,274)(1391,274)(1376,273)(1361,273)(1346,273)(1332,273)(1317,274)(1302,274)(1287,274)(1272,274)(1257,274)(1242,275)(1227,275)(1212,275)(1197,276)(1182,277)(1167,277)(1151,278)(1136,279)(1121,280)(1106,280)(1090,282)(1075,282)(1060,284)(1044,285)(1029,286)(1013,287)(998,289)(982,291)(967,293)(951,294)(936,297)(920,298)(904,301)(889,303)(873,307)(857,309)(841,313)
\thinlines \dashline[-10]{8}(841,313)(825,318)(810,323)(794,329)(778,338)
\thicklines \path(370,249)(2276,249)(2276,1718)(370,1718)(370,249)
\end{picture}
\caption{The allowed region(s) of $\sin2\phi_D $
using the experimental results in Table~\ref{CPdata}
assuming the direct CP asymmetries are zero.
The region shown by dotted lines is
coming from the ambiguity solution for $\sin2 \phi_1 $. }
\label{F1}
\end{center}
\end{wrapfigure}

Let us consider a special case, in which all the direct CP
asymmetries are exactly zero. In this case we can estimate the
allowed range of $\sin 2\phi_D $ (see Eq.~\ref{eq:Sf}) using the
experimental results in Table~\ref{CPdata}, as shown in
Fig.~\ref{F1}. Ignoring the region from the ambiguity solution for
$\phi_1$, $\sin 2\phi_D $ seems to lie away from zero, taking a
value around $-0.3$ . If this is true, such a large weak decay
phase must also affect other related decay modes. For example, if
we measure, {\it with enough precision}, the time-dependent CP
asymmetry of $B_s \to K^0\overline{K}^0 $, which is a pure $b \to
s$ QCD penguin process, then we could directly extract such a
large new weak phase. Similarly, using $B_s \rightarrow
\eta^\prime \pi^0 $, one can investigate existence of new weak
phase in EW penguin sector. Therefore, in order to find the origin
of new weak phase, precision-measurements of the related $B_s $
decays are  very important. Some of the corresponding $B_s$ decay
modes are listed in Table~\ref{Bd-Bs}. Because there is
essentially no weak phase in $B_s-\overline{B}_s$ mixing within
the SM, any sizable $S_f $ can directly indicate an existence of a
weak phase from NP. Moreover, by comparing these decay modes, one
may figure out the origin of the new CP phase. Even if some new
phases enter the mixing process of $B_s$, one can still probe them
by considering various modes simultaneously.

\begin{table}[h]
\begin{center}
\begin{tabular}{|c|c||c|c|}\hline
  & $S_f = \sin(2\phi_1 + 2\phi_D ) $ & $~~S_f = \sin2\phi_D~~ $ &
        origin of $\phi_D $\\
\hline
VP & $B_d\rightarrow \phi K^0 $ & $B_s\rightarrow \phi \eta^\prime $
        & QCD and EW Penguins
\\[2mm]
$\eta^\prime$ P  & $B_d\rightarrow \eta^\prime K^0 $ &
          $B_s\rightarrow \eta^\prime \pi^0 $  &  EW Penguin\\[2mm]
PP &  $B_d\rightarrow K^0 \pi^0 $ &
                 $ B_s\rightarrow K^0 \overline{K}^0$ & $b \to s$ QCD Penguin
                                \\[2mm]
  &   $ B_d \rightarrow K^0 \overline{K}^0 $          &  & $b \to d$ QCD
          Penguin \\
\hline
\end{tabular}
\caption{The $B_d$ and $B_s$ decay modes to VP, PP and $\eta'$P,
showing the origin of the weak decay phase $\phi_D$.}
\label{Bd-Bs}
\end{center}
\end{table}

If there exist such NP effects in $b\to s$ penguin process, they
might also appear in $b\to d$ penguin sector and so influence
relevant decay modes such as $B\rightarrow \pi\pi$ decays. This
may cause a serious concern for extracting $\phi_2$ from the
time-dependent CP asymmetry of $B\rightarrow \pi\pi$. One could
use $B_d \rightarrow K^0 \overline{K}^0$ mode to extract necessary
information about such effects in $b\to d$ penguin diagrams.
Time-dependent CP asymmetry of $B_d \rightarrow K^0
\overline{K}^0$ can also play an important role to understand the
difference of $S_f$'s from $B\rightarrow J/\psi K^0 $ and $b\to s$
penguin processes. In particular, this measurement will be useful
to clarify the relations among the branching fractions and CP
asymmetries of $B\rightarrow K\pi $ and $B\rightarrow \pi \pi $
modes which might contain a clue about the new weak phase in the
penguin contributions~\cite{BFRS,KP-PP-DCP,Lipkin,MY}.

For the purpose of exploring the origin of the discrepancies and
extracting possible NP contributions to CP-violating phases, this
paper is organized as follows: In Section 2, we classify the possible
cases for $S_f$'s of the relevant modes and describe each case in
detail.  Then we suggest how to use the $B_s$ decays to determine the
new CP-violating phase(s) in Section 3. In Section 4, our discussions
and conclusions are summarized.

\section{Classification}

In Table~\ref{CaseA-D}, we list four possibilities by classifying
the CP asymmetries of the three $b\to s$ penguin processes ($B \to
K \pi,~\phi K,~\eta' K$) compared to the SM prediction of
$B_d\rightarrow J/\psi K$. {\bf Case A} indicates that the origin
of new CP phase may reside in $b \to s \overline ss$ or singlet
QCD penguin sector, and so the effect should appear in $S_{\phi
K^0}$ and $S_{\eta^\prime K}$. If there is a new contribution in
color-suppressed tree diagram, its effects have to appear in
$S_{K\pi}$ and $S_{\eta^\prime K}$, which is {\bf Case B}. {\bf
Case C} is a rather accidental one because there is no common
factor to distinguish only $B_d\rightarrow \eta^\prime K $ from
the other two modes. Although it may not be impossible to realize,
it will be difficult to find a simple explanation. All these three
modes have contributions from QCD and EW penguins so that if any
sizable new weak phases exist in penguin sectors, the effect must
appear in all the modes. This corresponds to {\bf Case D}.

\begin{table}[ht]
\begin{center}
\begin{tabular}{|c|c|c|}\hline
{\bf Case}  & Condition & Possible Source of New Physics \\
\hline
{\bf A} & $S_{J/\psi K} \simeq  S_{K^0\pi^0}
                  \neq S_{\phi K} \simeq S_{\eta^\prime K} $ &
        $b \to s \overline s s$ process, singlet QCD penguin type
\\[2mm]
{\bf B}  & $S_{J/\psi K} \simeq  S_{\phi K^0}
                  \neq S_{K^0 \pi^0} \simeq S_{\eta^\prime K}$ &
          color suppressed tree, $b \to s \overline q q$ process \\[2mm]
{\bf C} &  $S_{J/\psi K} \simeq  S_{\eta^\prime K^0}
                  \neq S_{\phi K} \simeq S_{K^0 \pi^0 } $ &
                possibly accidental \\[2mm]
{\bf D} & $S_{J/\psi K} \neq  S_{K^0 \pi^0}
                  \simeq S_{\phi K} \simeq S_{\eta^\prime K} $
             & QCD penguin type or
               EW penguin type process  \\
\hline
\end{tabular}
\caption{Four possibilities in the CP asymmetries of the three
$b\to s$ penguin processes ($B \to K \pi,~\phi K,~\eta' K$),
compared to the SM prediction of $B_d\rightarrow J/\psi K$.  }
\label{CaseA-D}
\end{center}
\end{table}

Present experimental data gives $S_{\phi K} \simeq S_{\eta^\prime
K} \simeq S_{K^0 \pi^0}~\sim 0.48$, hence we may say that the
current situation is fairly close to {\bf Case D}. As can be seen
in Fig.~\ref{F1}, the current experimental results seem to
indicate that a sizable new CP phase $\phi_D$ is needed to explain
the discrepancy.  Since the weak phases from the $B-\overline{B}$
mixing are known to be all the same, the origin of the differences
must lie in the decay processes. On the other hand, the direct CP
asymmetries $A_f$'s are consistent with zero for all the modes. It
can be controlled by the strong phase difference. Now we consider
each case in detail.\\

{\bf Case A (with $\bf S_{J/\psi K} \simeq  S_{K^0\pi^0} \neq
S_{\phi K} \simeq S_{\eta^\prime K} $) }

In this case NP will appear only in $b \to s \overline s s$
type processes. We parametrize its contribution
as follows:
\bea
(P^s + S^s - \frac{1}{3}P^s_{EW} )V_{tb}^* V_{ts} \Rightarrow
A_{SM}^s
 ~( 1 + r e^{i\delta } e^{i\theta } ),
\eea where $A_{SM}$ is the SM prediction. And $r$ is the relative
ratio of summed NP contributions to the SM contribution,
$\delta $ is the resultant NP strong phase relative to the SM,
and $\theta$ is the relative CP phase
from NP. Strong phase $\delta $ depends on the decay process,
but weak phase $\theta $ depends only on interaction type. For simplicity,
we assume that the weak phase differences $\theta $ for $b \to s$
penguin type diagrams are same because these three modes
are almost pure penguin processes.

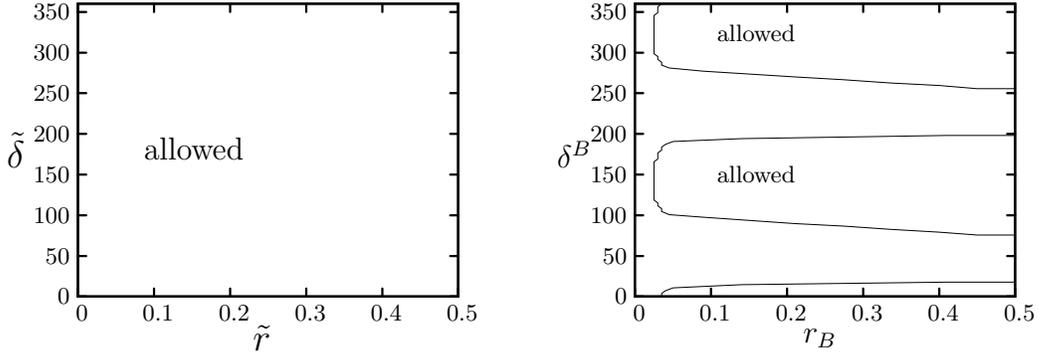
\begin{figure}[htbp]
\begin{center}
\begin{minipage}[l]{2.5in}
\setlength{\unitlength}{0.075450pt}
\begin{picture}(2400,1800)(0,0)
\footnotesize
\thicklines \path(370,249)(411,249)
\thicklines \path(2276,249)(2235,249)
\put(329,249){\makebox(0,0)[r]{ 0}}
\thicklines \path(370,453)(411,453)
\thicklines \path(2276,453)(2235,453)
\put(329,453){\makebox(0,0)[r]{ 50}}
\thicklines \path(370,657)(411,657)
\thicklines \path(2276,657)(2235,657)
\put(329,657){\makebox(0,0)[r]{ 100}}
\thicklines \path(370,861)(411,861)
\thicklines \path(2276,861)(2235,861)
\put(329,861){\makebox(0,0)[r]{ 150}}
\thicklines \path(370,1065)(411,1065)
\thicklines \path(2276,1065)(2235,1065)
\put(329,1065){\makebox(0,0)[r]{ 200}}
\thicklines \path(370,1269)(411,1269)
\thicklines \path(2276,1269)(2235,1269)
\put(329,1269){\makebox(0,0)[r]{ 250}}
\thicklines \path(370,1473)(411,1473)
\thicklines \path(2276,1473)(2235,1473)
\put(329,1473){\makebox(0,0)[r]{ 300}}
\thicklines \path(370,1677)(411,1677)
\thicklines \path(2276,1677)(2235,1677)
\put(329,1677){\makebox(0,0)[r]{ 350}}
\thicklines \path(370,249)(370,290)
\thicklines \path(370,1718)(370,1677)
\put(370,166){\makebox(0,0){ 0}}
\thicklines \path(751,249)(751,290)
\thicklines \path(751,1718)(751,1677)
\put(751,166){\makebox(0,0){ 0.1}}
\thicklines \path(1132,249)(1132,290)
\thicklines \path(1132,1718)(1132,1677)
\put(1132,166){\makebox(0,0){ 0.2}}
\thicklines \path(1514,249)(1514,290)
\thicklines \path(1514,1718)(1514,1677)
\put(1514,166){\makebox(0,0){ 0.3}}
\thicklines \path(1895,249)(1895,290)
\thicklines \path(1895,1718)(1895,1677)
\put(1895,166){\makebox(0,0){ 0.4}}
\thicklines \path(2276,249)(2276,290)
\thicklines \path(2276,1718)(2276,1677)
\put(2276,166){\makebox(0,0){ 0.5}}
\thicklines \path(370,249)(2276,249)(2276,1718)(370,1718)(370,249)
\put(12,983){\makebox(0,0)[l]{\Large $\tilde{\delta } $ }}
\put(1323,42){\makebox(0,0){\Large $\tilde{r} $  }}
\put(977,990){\makebox(0,0){\large allowed  }}
\thicklines \path(370,249)(2276,249)(2276,1718)(370,1718)(370,249)
\end{picture}
\end{minipage}
    \hspace*{8mm}
\begin{minipage}[r]{2.5in}
\setlength{\unitlength}{0.075450pt}
\begin{picture}(2400,1800)(0,0)
\footnotesize
\thicklines \path(370,249)(411,249)
\thicklines \path(2276,249)(2235,249)
\put(329,249){\makebox(0,0)[r]{ 0}}
\thicklines \path(370,453)(411,453)
\thicklines \path(2276,453)(2235,453)
\put(329,453){\makebox(0,0)[r]{ 50}}
\thicklines \path(370,657)(411,657)
\thicklines \path(2276,657)(2235,657)
\put(329,657){\makebox(0,0)[r]{ 100}}
\thicklines \path(370,861)(411,861)
\thicklines \path(2276,861)(2235,861)
\put(329,861){\makebox(0,0)[r]{ 150}}
\thicklines \path(370,1065)(411,1065)
\thicklines \path(2276,1065)(2235,1065)
\put(329,1065){\makebox(0,0)[r]{ 200}}
\thicklines \path(370,1269)(411,1269)
\thicklines \path(2276,1269)(2235,1269)
\put(329,1269){\makebox(0,0)[r]{ 250}}
\thicklines \path(370,1473)(411,1473)
\thicklines \path(2276,1473)(2235,1473)
\put(329,1473){\makebox(0,0)[r]{ 300}}
\thicklines \path(370,1677)(411,1677)
\thicklines \path(2276,1677)(2235,1677)
\put(329,1677){\makebox(0,0)[r]{ 350}}
\thicklines \path(370,249)(370,290)
\thicklines \path(370,1718)(370,1677)
\put(370,166){\makebox(0,0){ 0}}
\thicklines \path(751,249)(751,290)
\thicklines \path(751,1718)(751,1677)
\put(751,166){\makebox(0,0){ 0.1}}
\thicklines \path(1132,249)(1132,290)
\thicklines \path(1132,1718)(1132,1677)
\put(1132,166){\makebox(0,0){ 0.2}}
\thicklines \path(1514,249)(1514,290)
\thicklines \path(1514,1718)(1514,1677)
\put(1514,166){\makebox(0,0){ 0.3}}
\thicklines \path(1895,249)(1895,290)
\thicklines \path(1895,1718)(1895,1677)
\put(1895,166){\makebox(0,0){ 0.4}}
\thicklines \path(2276,249)(2276,290)
\thicklines \path(2276,1718)(2276,1677)
\put(2276,166){\makebox(0,0){ 0.5}}
\thicklines \path(370,249)(2276,249)(2276,1718)(370,1718)(370,249)
\put(-20,963){\makebox(0,0)[l]{\large $\delta^B $ }}
\put(1323,42){\makebox(0,0){\large $ r_B $  }}
\put(977,1570){\makebox(0,0){ allowed  }}
\put(977,865){\makebox(0,0){ allowed  }}
\thinlines \path(503,249)(503,264)(522,278)(561,293)(904,308)(1933,322)(2276,322)
\thinlines \path(2276,557)(2085,557)(2085,557)(1895,572)(1647,587)(1418,602)(1171,616)(942,631)(713,646)(542,660)(503,675)(503,690)(484,704)(484,719)(465,734)(465,748)(465,763)(465,778)(465,793)(465,807)(465,822)(465,837)(465,851)(465,866)(465,881)(465,895)(465,910)(465,925)(484,939)(484,954)(484,969)(503,984)(503,998)(522,1013)(561,1028)(904,1042)(1933,1057)(2276,1057)
\thinlines \path(2276,1292)(2085,1292)(2085,1292)(1895,1307)(1647,1321)(1418,1336)(1171,1351)(942,1365)(713,1380)(542,1395)(503,1410)(503,1424)(484,1439)(484,1454)(465,1468)(465,1483)(465,1498)(465,1512)(465,1527)(465,1542)(465,1556)(465,1571)(465,1586)(465,1600)(465,1615)(465,1630)(465,1645)(465,1659)(484,1674)(484,1689)(484,1703)(503,1718)(2276,1718)
\thicklines \path(370,249)(2276,249)(2276,1718)(370,1718)(370,249)
\end{picture}
\end{minipage}
\caption{The allowed region of the parameters $\tilde{\delta} $ and
  $\tilde{r}$ for $B\rightarrow \phi K^0$ decay (Left) and $\delta^B $
  and $r_{_B}$ for $B\rightarrow \eta^\prime K_S$ decay (Right) for
  {\bf Case A}.  Here $\theta $ is assumed as a free parameter.}
    \label{fig:2}
\end{center}
\end{figure}

Neglecting the $C$ term, one can
obtain \bea A(B\rightarrow \phi K ) &=& \tilde{A}_{SM}^s
             ( 1 + \tilde{r} e^{i\tilde{\delta } } e^{i\theta } ), \\
A(B\rightarrow \eta^\prime K^0 ) &=& \frac{1}{\sqrt{6}}\left[
      A_{SM}
 + 2 {A}_{SM}^s
             ( 1 + r e^{i\delta } e^{i\theta } ) \right] \nn \\
    &\equiv & {B}_{SM}
             ( 1 + r_B e^{i\delta^B } e^{i\theta } ),
\eea where $\tilde{\delta }$ and $\delta $ are the relative strong
phases and $\theta $ is a CP phase from NP. Here we assume that
the same new weak phase enters the two modes because its origin is
the same type of penguin diagram. In the $B_d\rightarrow
\eta^\prime K^0 $ decay, the re-defined parameters are
\bea
B_{SM} &=& \frac{1}{\sqrt{6}}(2 A_{SM}^s + A_{SM} ) , \\
B_{SM} r_B e^{i\delta^B } &=& \frac{2}{\sqrt{6}} A_{SM}^s r
e^{i\delta },
\eea
where there is no NP contribution in
$B\rightarrow K^0\pi^0$, because it is not $b \to s\overline{s}s$
process. Assuming that there is no direct CP violation in the SM
and so $|A_{SM}| = |\overline{A}_{SM}|$, then the measurements are
expressed as
\bea Br_{\phi K} &\propto & |\tilde{A}_{SM}^s|^2
              (1 + {\tilde{r}^2} + 2 \tilde{r} \cos\tilde{\delta }
                      \cos\theta ), \\
A_{\phi K} &\equiv& - \frac{|A|^2 - |\overline{A}|^2 }{|A|^2 + |\overline{A}|^2}
             = \frac{2 \tilde{r} \sin\tilde{\delta }\sin\theta }
                      {1 + \tilde{r}^2 + 2 \tilde{r} \cos\tilde{\delta
              }
                       \cos\theta }, \\
S_{\phi K} &\equiv&
            \frac{2 Im(e^{-2i\phi_1} A^* \overline{A})}{|A|^2 + |\overline{A}|^2}
              = \frac{\sin{2 \phi_1 } + \tilde{r}^2 \sin2(\phi_1+\theta)
                      + 2 \tilde{r} \cos\tilde{\delta}
                                     \sin(2 \phi_1 + \theta ) }
                    { 1 + \tilde{r}^2 + 2 \tilde{r} \cos\tilde{\delta
              }
                    \cos\theta }.
\eea
The weak phase $\phi_1$ from the $B^0_d-\overline{B}^0_d$ mixing is
supposed to be extracted from $B\rightarrow J/\psi K_s $ at good
accuracy.  Similarly, the expressions for $B\rightarrow \eta^\prime
K^0 $ are obtained by
\bea
Br_{\eta^\prime K} &\propto & |{B}_{SM}|^2
              (1 + r_B^2 + 2 r_B \cos\delta^B \cos\theta ), \\
A_{\eta^\prime K} &=& \frac{2 r_B \sin\delta^B \sin\theta }
                      {1 + r_B^2 + 2 r_B \cos\delta^B \cos\theta }, \\
S_{\eta^\prime K} &=& \frac{\sin{2 \phi_1 } + r_B^2 \sin2(\phi_1+\theta)
                      + 2 r_B \cos\delta^B \sin(2 \phi_1 + \theta ) }
                    { 1 + r_B^2 + 2 r_B \cos\delta^B \cos\theta }.
\eea
In Fig.~\ref{fig:2}, we show the allowed parameter space by using the
current experimental constraints. According to our estimate,
$\delta^B$ should be around $0^\circ $ or $180^\circ $ because the
experimental data of $A_{\eta^\prime K}$ is almost zero and the error
is very small.

\begin{figure}[htb]
\begin{center}
\begin{minipage}[l]{2.5in}
\setlength{\unitlength}{0.075450pt}
\begin{picture}(2400,1800)(0,0)
\footnotesize
\thicklines \path(370,249)(411,249)
\thicklines \path(2276,249)(2235,249)
\put(329,249){\makebox(0,0)[r]{ 0}}
\thicklines \path(370,453)(411,453)
\thicklines \path(2276,453)(2235,453)
\put(329,453){\makebox(0,0)[r]{ 50}}
\thicklines \path(370,657)(411,657)
\thicklines \path(2276,657)(2235,657)
\put(329,657){\makebox(0,0)[r]{ 100}}
\thicklines \path(370,861)(411,861)
\thicklines \path(2276,861)(2235,861)
\put(329,861){\makebox(0,0)[r]{ 150}}
\thicklines \path(370,1065)(411,1065)
\thicklines \path(2276,1065)(2235,1065)
\put(329,1065){\makebox(0,0)[r]{ 200}}
\thicklines \path(370,1269)(411,1269)
\thicklines \path(2276,1269)(2235,1269)
\put(329,1269){\makebox(0,0)[r]{ 250}}
\thicklines \path(370,1473)(411,1473)
\thicklines \path(2276,1473)(2235,1473)
\put(329,1473){\makebox(0,0)[r]{ 300}}
\thicklines \path(370,1677)(411,1677)
\thicklines \path(2276,1677)(2235,1677)
\put(329,1677){\makebox(0,0)[r]{ 350}}
\thicklines \path(370,249)(370,290)
\thicklines \path(370,1718)(370,1677)
\put(370,166){\makebox(0,0){ 0}}
\thicklines \path(751,249)(751,290)
\thicklines \path(751,1718)(751,1677)
\put(751,166){\makebox(0,0){ 0.1}}
\thicklines \path(1132,249)(1132,290)
\thicklines \path(1132,1718)(1132,1677)
\put(1132,166){\makebox(0,0){ 0.2}}
\thicklines \path(1514,249)(1514,290)
\thicklines \path(1514,1718)(1514,1677)
\put(1514,166){\makebox(0,0){ 0.3}}
\thicklines \path(1895,249)(1895,290)
\thicklines \path(1895,1718)(1895,1677)
\put(1895,166){\makebox(0,0){ 0.4}}
\thicklines \path(2276,249)(2276,290)
\thicklines \path(2276,1718)(2276,1677)
\put(2276,166){\makebox(0,0){ 0.5}}
\thicklines \path(370,249)(2276,249)(2276,1718)(370,1718)(370,249)
\put(-22,983){\makebox(0,0)[l]{\shortstack{{\Large $\theta $}}}}
\put(1323,42){\makebox(0,0){\Large $ r $ }}
\put(977,1370){\makebox(0,0){ allowed  }}
\put(977,670){\makebox(0,0){ allowed  }}
\thinlines \path(2276,264)(885,264)(885,264)(637,278)(561,293)(522,308)(484,322)(465,337)(465,352)(446,367)(446,381)(446,396)(427,411)(427,425)(427,440)(427,455)(427,469)(427,484)(427,499)(427,513)(408,528)(408,543)(408,557)(408,572)(408,587)(408,602)(408,616)(408,631)(408,646)(408,660)(408,675)(408,690)(408,704)(427,719)(427,734)(427,748)(427,763)(427,778)(427,793)(427,807)(427,822)(446,837)(446,851)(446,866)(465,881)(484,895)(484,910)(522,925)(561,939)(675,954)(980,969)(2276,969)
\thinlines \path(2276,998)(885,998)(885,998)(637,1013)(561,1028)(522,1042)(484,1057)(465,1072)(465,1086)(446,1101)(446,1116)(446,1130)(427,1145)(427,1160)(427,1174)(427,1189)(427,1204)(427,1219)(427,1233)(427,1248)(408,1263)(408,1277)(408,1292)(408,1307)(408,1321)(408,1336)(408,1351)(408,1365)(408,1380)(408,1395)(408,1410)(408,1424)(408,1439)(427,1454)(427,1468)(427,1483)(427,1498)(427,1512)(427,1527)(427,1542)(427,1556)(446,1571)(446,1586)(446,1600)(465,1615)(484,1630)(484,1645)(522,1659)(561,1674)(675,1689)(980,1703)(2276,1703)
\thicklines \path(370,249)(2276,249)(2276,1718)(370,1718)(370,249)
\end{picture}
\end{minipage}
    \hspace*{8mm}
\begin{minipage}[r]{2.5in}
\setlength{\unitlength}{0.075450pt}
\begin{picture}(2400,1800)(0,0)
\footnotesize
\thicklines \path(370,249)(411,249)
\thicklines \path(2276,249)(2235,249)
\put(329,249){\makebox(0,0)[r]{ 0}}
\thicklines \path(370,453)(411,453)
\thicklines \path(2276,453)(2235,453)
\put(329,453){\makebox(0,0)[r]{ 50}}
\thicklines \path(370,657)(411,657)
\thicklines \path(2276,657)(2235,657)
\put(329,657){\makebox(0,0)[r]{ 100}}
\thicklines \path(370,861)(411,861)
\thicklines \path(2276,861)(2235,861)
\put(329,861){\makebox(0,0)[r]{ 150}}
\thicklines \path(370,1065)(411,1065)
\thicklines \path(2276,1065)(2235,1065)
\put(329,1065){\makebox(0,0)[r]{ 200}}
\thicklines \path(370,1269)(411,1269)
\thicklines \path(2276,1269)(2235,1269)
\put(329,1269){\makebox(0,0)[r]{ 250}}
\thicklines \path(370,1473)(411,1473)
\thicklines \path(2276,1473)(2235,1473)
\put(329,1473){\makebox(0,0)[r]{ 300}}
\thicklines \path(370,1677)(411,1677)
\thicklines \path(2276,1677)(2235,1677)
\put(329,1677){\makebox(0,0)[r]{ 350}}
\thicklines \path(370,249)(370,290)
\thicklines \path(370,1718)(370,1677)
\put(370,166){\makebox(0,0){ 0}}
\thicklines \path(751,249)(751,290)
\thicklines \path(751,1718)(751,1677)
\put(751,166){\makebox(0,0){ 0.1}}
\thicklines \path(1132,249)(1132,290)
\thicklines \path(1132,1718)(1132,1677)
\put(1132,166){\makebox(0,0){ 0.2}}
\thicklines \path(1514,249)(1514,290)
\thicklines \path(1514,1718)(1514,1677)
\put(1514,166){\makebox(0,0){ 0.3}}
\thicklines \path(1895,249)(1895,290)
\thicklines \path(1895,1718)(1895,1677)
\put(1895,166){\makebox(0,0){ 0.4}}
\thicklines \path(2276,249)(2276,290)
\thicklines \path(2276,1718)(2276,1677)
\put(2276,166){\makebox(0,0){ 0.5}}
\thicklines \path(370,249)(2276,249)(2276,1718)(370,1718)(370,249)
\put(-22,983){\makebox(0,0)[l]{\large $\delta $}}
\put(1323,42){\makebox(0,0){\Large $ r $ }}
\put(977,1500){\makebox(0,0){ allowed  }}
\put(977,800){\makebox(0,0){ allowed  }}
\thinlines \path(408,249)(408,264)(408,278)(427,293)(427,308)(427,322)(446,337)(465,352)(580,367)(1171,381)(1838,396)(2276,396)
\thinlines \path(2276,440)(1990,440)(1990,440)(1685,455)(1342,469)(980,484)(637,499)(465,513)(446,528)(427,543)(427,557)(427,572)(408,587)(408,602)(408,616)(408,631)(408,646)(408,660)(408,675)(408,690)(408,704)(408,719)(408,734)(408,748)(408,763)(408,778)(408,793)(408,807)(408,822)(408,837)(408,851)(408,866)(408,881)(408,895)(408,910)(408,925)(408,939)(408,954)(408,969)(408,984)(408,998)(408,1013)(427,1028)(427,1042)(427,1057)(446,1072)(465,1086)(580,1101)(1171,1116)(1838,1130)(2276,1130)
\thinlines \path(2276,1174)(1990,1174)(1990,1174)(1685,1189)(1342,1204)(980,1219)(637,1233)(465,1248)(446,1263)(427,1277)(427,1292)(427,1307)(408,1321)(408,1336)(408,1351)(408,1365)(408,1380)(408,1395)(408,1410)(408,1424)(408,1439)(408,1454)(408,1468)(408,1483)(408,1498)(408,1512)(408,1527)(408,1542)(408,1556)(408,1571)(408,1586)(408,1600)(408,1615)(408,1630)(408,1645)(408,1659)(408,1674)(408,1689)(408,1703)(408,1718)
\thicklines \path(370,249)(2276,249)(2276,1718)(370,1718)(370,249)
\end{picture}
\end{minipage}
\caption{The $1 \sigma $ allowed region for 3 parameters determined
  from the 4 measurements for {\bf Case A}. 
}
\label{fig:3}
\end{center}
\end{figure}
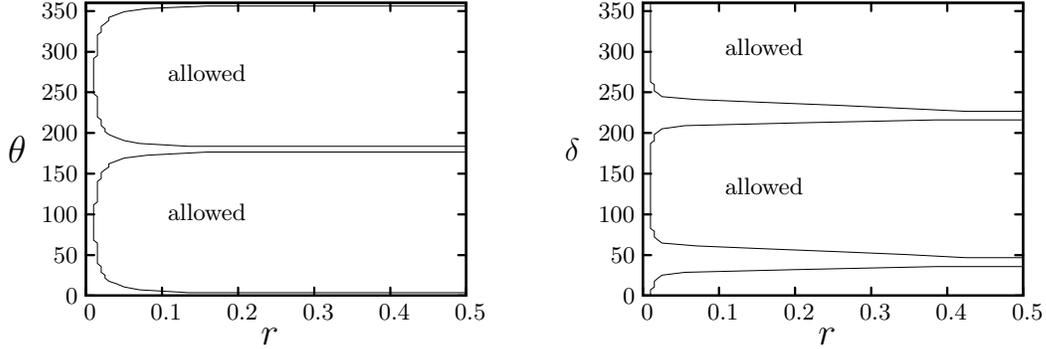

If we consider exact SU(3) flavor symmetry, we can obtain several relations
among the parameters.  Under the special condition,
the strong phases should be same and also
$|A_{SM}|=|A_{SM}^s|$ so that
\bea
\delta &=& \delta^B,  \\
\frac{r_B}{r} &=& \frac{2 A_{SM}^s }{2 A_{SM}^s + A_{SM}} =
\frac{2}{3}.
\eea
For simplicity,
though it is not quite precise due to non-negligible
SU(3) breaking effects and possible final state re-scattering
effects\footnote{
Here we are only trying to find a method to classify
roughly the dependence of the parameters from NP effects, so that we
neglect all those small  effects.},
we consider the following relations to get the rough
estimation of the allowed region by reducing the number of parameters,
\bea
r &=& \tilde{r}, \\
\delta &=& \tilde{\delta } = \delta^B .
\eea
Using these relations, we can extract the 3 parameters, $r,\delta$ and
$\theta $ from the 4 measurements. The solutions (allowed regions) are
shown in Fig.~\ref{fig:3}. These figures tell us we can not still so
strictly constrain the parameter regions by the present experimental
data but one can roughly see the dependence. Here we have to note
that, as we mentioned before, this analysis has been made upon somewhat
rough assumptions, therefore we may need more thorough considerations.
As mentioned earlier, the penguin diagrams may be including different
contributions for $B\rightarrow VP$ and $B\rightarrow PP$ so that we
can not describe them with the same parameters.  However, we find that
these estimates are consistent with the allowed region
for
$\sin2\phi_D $ in Fig.~\ref{F1}, where $\sin2\phi_D $ lies around
$-0.3$.  Considering the relation between $\sin\phi_D$ and
$\sin\theta$
for $\delta \sim0^\circ $ (no direct CP asymmetry case), which is
derived from $ 2\phi_D = Arg\frac{A(B^0 \rightarrow
f_{CP})}{A(\bar{B}^0\rightarrow f_{CP})} $,
\bea
\sin 2\phi_D = - 2 \frac{r}{1+r^2 + 2 r \cos\theta}
              \sin\theta (1 + r \cos\theta ),
\eea
one can see that the estimates, $\sin\theta \sim 1$ and $r\sim 0.15 $
are consistent with those in Fig.~\ref{F1}. Thus in the sense seeing
the rough estimation, one can find the dependence among the unknown
parameters and may get some hints of new physics
if we can find more accurate experimental data under this Case A\footnote{
Note that we cannot directly compare the results within
the allowed regions of Fig.~\ref{F1}.
Here we use the comparison of Fig.~\ref{F1}
as a reference point to check whether our rough estimate
can lead comparatively correct answer.
}.
\\

{\bf Case B (with $\bf S_{J/\psi K} \simeq  S_{\phi K^0}
\neq S_{K^0 \pi^0} \simeq S_{\eta^\prime K}$) }

In this case the possible source of NP could be in color-suppressed
tree diagram or $b \to s \overline q q$ penguin sector. Although it
may cause another difficulties, large
color-suppressed tree contribution may come from NP or even in the SM
due to our misunderstanding about how to estimate it. Actually, in
recent several works~\cite{KP-PP-DCP,recent-PP-KP} on $B\rightarrow
K\pi $, this possibility has been considered to explain the difference
in the direct CP asymmetries of $B\rightarrow K^+\pi^- $ and
$B\rightarrow K^+\pi^0 $. The large $C$ contribution may also be
useful to explain the discrepancies in the branching ratios of
$B\rightarrow \pi^0 \pi^0 $, and etc.
Therefore, we divide this case into two parts to discuss the case within the
SM with unexpectedly large $C$, and the case with new weak phase from NP.
\\

{\bf Case B-1 : The weak phase in the color-suppressed tree  is only from
CKM phase.}

Considering penguin and tree type contributions separately,
the decay amplitudes can be written as follows:
\bea
A(B\rightarrow \eta^\prime K^0 ) &=& {A}_{P} (1 - {r}
                                               e^{i{\delta }}
                                                          e^{i\phi_3}
                                                          ), \\
A(B\rightarrow \pi^0 K^0 ) &=& {A}^w_{P} (1 +{w}
                                               e^{i{\delta }}
                                               e^{i\phi_3}
                                                          ),
\eea
where $\delta$ is a strong phase difference and $r$ and $w$ represent
the relative ratios of the color-suppressed
tree to the penguin contributions.
Here we assume that the strong phase difference be the same in both modes.
There are 3 unknown parameters and 1 weak phase for 4
measurements given by
\bea
A_{\eta^\prime K} &=& - \frac{2 r \sin\delta \sin\phi_3 }
                      {1 + r^2 - 2 r \cos\delta \cos\phi_3 }, \\
S_{\eta^\prime K} &=& \frac{\sin{2 \phi_1 } + r^2 \sin2(\phi_1+\phi_3)
                      - 2 r \cos\delta \sin(2 \phi_1 + \phi_3 ) }
                    { 1 + r^2 - 2 r \cos\delta \cos\phi_3 },
                      \\[1mm]
A_{\pi^0 K} &=& \frac{2 w \sin\delta \sin\phi_3 }
                      {1 + w^2 + 2 w \cos\delta \cos\phi_3 }, \\
S_{\pi^0 K} &=& \frac{\sin{2 \phi_1 } + w^2 \sin2(\phi_1+\phi_3)
                      + 2 w \cos\delta \sin(2 \phi_1 + \phi_3 ) }
                    { 1 + w^2 + 2 w \cos\delta \cos\phi_3 }.
\eea
Considering $\sin2\phi_1\sim 0.69$, there should be some destructive
contributions in the numerator of the above $S_f$ to satisfy the
current experimental results.  It could be done by requiring $90^\circ
< (\phi_1+\phi_3) < 180^\circ $ or controlling by the strong phase
difference $\delta$ in the third term. However, the explanation by the
strong phase difference is somewhat problematic, because the signs of
the terms with $\cos\delta$ are different in the two decay modes so
that its contribution will be always opposite.  Requiring $90^\circ <
(\phi_1+\phi_3) < 180^\circ $ seems to be rather difficult because it
needs large $\phi_3 $ compared to the current CKM bounds\footnote{
The recent estimates
are $\phi_3 = (59.8^{+4.9}_{-4.1}){}^\circ $ by CKM fitter Group\cite{CKMfit}
and $\phi_3 = (61.3\pm4.5){}^\circ $ by UT fit group\cite{UTfit},
which means that $\phi_1+\phi_3 < 90^\circ $ is favored.}.
And then one may consider a possibility that the strong phases are not the
same in the two decay modes. But this again, we will need at least one
large strong phase to reduce both $S_f$'s.  In addition, the sign of
the direct CP asymmetries could also give us some information about
the strong phases, but the origin of such large differences will
remain as a problem.

In fact,
if we assume that the color-suppressed tree contribution ($C$)
is large~\cite{KP-PP-DCP} to explain several discrepancies
in $B\rightarrow K\pi $ and $\pi \pi $
decays\cite{BFRS,Lipkin,MY,recent-PP-KP},
it may also cause
another discrepancy between $B\rightarrow \phi K $ and the other
$b \to s$ penguin decay modes. As can be seen in
Table~\ref{CPdata}, there is no such indication from experimental
data. In contrast, if $C$ is
negligible, the direct CP violations will be almost zero and the
indirect CP asymmetries are predicted as $S_f = \sin2\phi_1$ for
all the three modes within the SM. In fact, considering the CKM
factors together, the color-suppressed tree contribution, $C
V_{ub}^* V_{us}$, is indeed very small compared with QCD penguin
contribution.
Therefore, in order to explain the discrepancy between CP asymmetries of
$B\rightarrow J/\psi K$ and the other modes,
it will be more appealing to have NP effects in $b\to s$
penguin processes than requiring large $C$ contributions within the SM.

As a temporary summery, recent experimental data seem not to prefer the explanation with
only large $C$ contribution within the SM. To explain them by
using large $C$ scenario, we need at least one more parameter like a
new weak phase difference. In {\bf Case B-2}, we will consider whether
the new weak phase can help to explain this scenario.\\

{\bf Case B-2 : Case B-1 plus a new physics weak phase.}

As a more general case, we now consider the case with the
new physics weak phase,
denoted by $\theta$. We can consider that this case may include both
cases: (a) Penguin sector has a new physics contribution with
new weak phase and (b) Color-suppressed tree sector has a new weak phase.
However, because $C$ is only a tree contribution so that the case (b)
may not be
acceptable that it has new physics contribution with a new phase
difference. Therefore, here we consider the case that the penguin has new physics
contribution with a new weak phase as well as with a possibility that
the magnitude of $C$ may be also larger than ordinary estimate
from the SM\footnote{
The only difference from the previous case is the new physics phase $\theta$.
Hence, we may think that {\bf Case B-1} is a special case of {\bf Case
B-2}.}.

Then, similar to the previous case, the four CP
asymmetries are given by
\bea
A_{\eta^\prime K} &=& - \frac{2 r \sin\delta \sin\theta }
                      {1 + r^2 - 2 r \cos\delta \cos\theta }, \\
S_{\eta^\prime K} &=& \frac{\sin{2 \phi_1 } + r^2 \sin2(\phi_1+\theta)
                      - 2 r \cos\delta \sin(2 \phi_1 + \theta ) }
                    { 1 + r^2 - 2 r \cos\delta \cos\theta },
                      \\[1mm]
A_{\pi^0 K} &=& \frac{2 w \sin\delta \sin\theta }
                      {1 + w^2 + 2 w \cos\delta \cos\theta }, \\
S_{\pi^0 K} &=& \frac{\sin{2 \phi_1 } + w^2 \sin2(\phi_1+\theta)
                      + 2 w \cos\delta \sin(2 \phi_1 + \theta ) }
                    { 1 + w^2 + 2 w \cos\delta \cos\theta }.
\eea


\begin{wrapfigure}{r}{7.5cm}
\begin{center}
\setlength{\unitlength}{0.088450pt}
\begin{picture}(2400,1800)(0,0)
\footnotesize
\thicklines \path(370,249)(411,249)
\thicklines \path(2276,249)(2235,249)
\put(329,249){\makebox(0,0)[r]{ 0}}
\thicklines \path(370,543)(411,543)
\thicklines \path(2276,543)(2235,543)
\put(329,543){\makebox(0,0)[r]{ 0.2}}
\thicklines \path(370,837)(411,837)
\thicklines \path(2276,837)(2235,837)
\put(329,837){\makebox(0,0)[r]{ 0.4}}
\thicklines \path(370,1130)(411,1130)
\thicklines \path(2276,1130)(2235,1130)
\put(329,1130){\makebox(0,0)[r]{ 0.6}}
\thicklines \path(370,1424)(411,1424)
\thicklines \path(2276,1424)(2235,1424)
\put(329,1424){\makebox(0,0)[r]{ 0.8}}
\thicklines \path(370,1718)(411,1718)
\thicklines \path(2276,1718)(2235,1718)
\put(329,1718){\makebox(0,0)[r]{ 1}}
\thicklines \path(370,249)(370,290)
\thicklines \path(370,1718)(370,1677)
\put(370,166){\makebox(0,0){ 0}}
\thicklines \path(688,249)(688,290)
\thicklines \path(688,1718)(688,1677)
\put(688,166){\makebox(0,0){ 0.05}}
\thicklines \path(1005,249)(1005,290)
\thicklines \path(1005,1718)(1005,1677)
\put(1005,166){\makebox(0,0){ 0.1}}
\thicklines \path(1323,249)(1323,290)
\thicklines \path(1323,1718)(1323,1677)
\put(1323,166){\makebox(0,0){ 0.15}}
\thicklines \path(1641,249)(1641,290)
\thicklines \path(1641,1718)(1641,1677)
\put(1641,166){\makebox(0,0){ 0.2}}
\thicklines \path(1958,249)(1958,290)
\thicklines \path(1958,1718)(1958,1677)
\put(1958,166){\makebox(0,0){ 0.25}}
\thicklines \path(2276,249)(2276,290)
\thicklines \path(2276,1718)(2276,1677)
\put(2276,166){\makebox(0,0){ 0.3}}
\thicklines \path(370,249)(2276,249)(2276,1718)(370,1718)(370,249)
\put(770,1460){\makebox(0,0)[l]{$ \theta = 80^\circ $  }}
\put(1110,1300){\makebox(0,0)[l]{$90^\circ $  }}
\put(1400,1300){\makebox(0,0)[l]{$95^\circ $  }}
\put(1450,730){\makebox(0,0)[l]{$100^\circ $  }}
\put(1950,750){\makebox(0,0)[l]{$110^\circ $  }}
\put(-22,983){\makebox(0,0)[l]{\large $ S_{\eta^\prime K }$ }}
\put(1323,42){\makebox(0,0){ \Large $r$  }}
\thinlines \path(370,1255)(370,1255)(402,1249)(434,1242)(465,1236)(497,1230)(529,1223)(561,1217)(592,1210)(624,1203)(656,1197)(688,1190)(719,1183)(751,1176)(783,1169)(815,1161)(847,1154)(878,1147)(910,1139)(942,1351)(974,1366)(1005,1378)(1037,1391)(1069,1402)(1101,1413)(1132,1423)(1164,1434)(1196,1442)(1228,1453)(1259,1461)(1291,1470)(1323,1478)(1355,1486)(1387,1494)(1418,1502)(1450,1510)(1482,1516)(1514,1523)(1545,1531)(1577,1536)(1609,1544)(1641,1549)(1672,1556)(1704,1561)(1736,1568)(1768,1573)(1800,1580)(1831,1584)(1863,1589)(1895,1595)(1927,1600)
\thinlines \path(1927,1600)(1958,1606)(1990,1610)(2022,1614)(2054,1618)(2085,1624)(2117,1627)(2149,1631)(2181,1634)(2212,1640)(2244,1643)(2276,1646)(2276,1646)
\thinlines \path(370,1255)(370,1255)(402,1246)(434,1238)(465,1229)(497,1221)(529,1212)(561,1203)(592,1194)(624,1185)(656,1176)(688,1167)(719,1157)(751,1148)(783,1138)(815,1129)(847,1119)(878,1109)(910,1100)(942,1090)(974,1080)(1005,1070)(1037,1060)(1069,1050)(1101,1420)(1132,1431)(1164,1442)(1196,1451)(1228,1462)(1259,1471)(1291,1479)(1323,1488)(1355,1496)(1387,1505)(1418,1513)(1450,1519)(1482,1527)(1514,1535)(1545,1541)(1577,1549)(1609,1557)(1641,1562)(1672,1569)(1704,1574)(1736,1582)(1768,1587)(1800,1591)(1831,1598)(1863,1603)(1895,1610)(1927,1614)
\thinlines \path(1927,1614)(1958,1618)(1990,1625)(2022,1629)(2054,1632)(2085,1636)(2117,1642)(2149,1646)(2181,1649)(2212,1652)(2244,1658)(2276,1660)(2276,1660)
\thinlines \path(370,1255)(370,1255)(402,1245)(434,1236)(465,1226)(497,1217)(529,1207)(561,1197)(592,1187)(624,1177)(656,1167)(688,1157)(719,1147)(751,1136)(783,1126)(815,1115)(847,1105)(878,1094)(910,1084)(942,1073)(974,1062)(1005,1051)(1037,1040)(1069,1029)(1101,1018)(1132,1007)(1164,996)(1196,985)(1228,974)(1259,963)(1291,951)(1323,940)(1355,1499)(1387,1505)(1418,1514)(1450,1522)(1482,1531)(1514,1536)(1545,1545)(1577,1553)(1609,1558)(1641,1566)(1672,1571)(1704,1579)(1736,1584)(1768,1591)(1800,1596)(1831,1601)(1863,1608)(1895,1612)(1927,1619)
\thinlines \path(1927,1619)(1958,1623)(1990,1627)(2022,1631)(2054,1638)(2085,1641)(2117,1645)(2149,1651)(2181,1654)(2212,1657)(2244,1660)(2276,1663)(2276,1663)
\thinlines \path(370,1255)(370,1255)(402,1244)(434,1234)(465,1224)(497,1213)(529,1203)(561,1192)(592,1182)(624,1171)(656,1160)(688,1149)(719,1139)(751,1128)(783,1117)(815,1106)(847,1094)(878,1083)(910,1072)(942,1061)(974,1049)(1005,1038)(1037,1027)(1069,1015)(1101,1004)(1132,992)(1164,981)(1196,969)(1228,957)(1259,946)(1291,934)(1323,922)(1355,911)(1387,899)(1418,887)(1450,875)(1482,864)(1514,852)(1545,840)(1577,828)(1609,816)(1641,804)(1672,793)(1704,781)(1736,769)(1768,757)(1800,745)(1831,733)(1863,721)(1895,710)(1927,698)
\thinlines \path(1927,698)(1958,686)(1990,674)(2022,662)(2054,650)(2085,639)(2117,627)(2149,615)(2181,603)(2212,592)(2244,580)(2276,568)(2276,568)
\thinlines \path(370,1255)(370,1255)(402,1245)(434,1234)(465,1224)(497,1214)(529,1204)(561,1193)(592,1183)(624,1173)(656,1162)(688,1152)(719,1141)(751,1131)(783,1120)(815,1109)(847,1099)(878,1088)(910,1077)(942,1067)(974,1056)(1005,1045)(1037,1034)(1069,1024)(1101,1013)(1132,1002)(1164,991)(1196,980)(1228,970)(1259,959)(1291,948)(1323,937)(1355,926)(1387,915)(1418,904)(1450,894)(1482,883)(1514,872)(1545,861)(1577,850)(1609,840)(1641,829)(1672,818)(1704,807)(1736,797)(1768,786)(1800,775)(1831,764)(1863,754)(1895,743)(1927,732)
\thinlines \path(1927,732)(1958,722)(1990,711)(2022,701)(2054,690)(2085,680)(2117,669)(2149,659)(2181,648)(2212,638)(2244,628)(2276,617)(2276,617)
\thinlines \dashline[-10]{12}(370,1116)(370,1116)(389,1116)(409,1116)(428,1116)(447,1116)(466,1116)(486,1116)(505,1116)(524,1116)(543,1116)(563,1116)(582,1116)(601,1116)(620,1116)(640,1116)(659,1116)(678,1116)(697,1116)(717,1116)(736,1116)(755,1116)(774,1116)(794,1116)(813,1116)(832,1116)(851,1116)(871,1116)(890,1116)(909,1116)(928,1116)(948,1116)(967,1116)(986,1116)(1005,1116)(1025,1116)(1044,1116)(1063,1116)(1082,1116)(1102,1116)(1121,1116)(1140,1116)(1159,1116)(1179,1116)(1198,1116)(1217,1116)(1236,1116)(1256,1116)(1275,1116)(1294,1116)(1313,1116)
\thinlines \dashline[-10]{12}(1313,1116)(1333,1116)(1352,1116)(1371,1116)(1390,1116)(1410,1116)(1429,1116)(1448,1116)(1467,1116)(1487,1116)(1506,1116)(1525,1116)(1544,1116)(1564,1116)(1583,1116)(1602,1116)(1621,1116)(1641,1116)(1660,1116)(1679,1116)(1698,1116)(1718,1116)(1737,1116)(1756,1116)(1775,1116)(1795,1116)(1814,1116)(1833,1116)(1852,1116)(1872,1116)(1891,1116)(1910,1116)(1929,1116)(1949,1116)(1968,1116)(1987,1116)(2006,1116)(2026,1116)(2045,1116)(2064,1116)(2083,1116)(2103,1116)(2122,1116)(2141,1116)(2160,1116)(2180,1116)(2199,1116)(2218,1116)(2237,1116)(2257,1116)(2276,1116)
\thinlines \dashline[-10]{12}(370,851)(370,851)(389,851)(409,851)(428,851)(447,851)(466,851)(486,851)(505,851)(524,851)(543,851)(563,851)(582,851)(601,851)(620,851)(640,851)(659,851)(678,851)(697,851)(717,851)(736,851)(755,851)(774,851)(794,851)(813,851)(832,851)(851,851)(871,851)(890,851)(909,851)(928,851)(948,851)(967,851)(986,851)(1005,851)(1025,851)(1044,851)(1063,851)(1082,851)(1102,851)(1121,851)(1140,851)(1159,851)(1179,851)(1198,851)(1217,851)(1236,851)(1256,851)(1275,851)(1294,851)(1313,851)
\thinlines \dashline[-10]{12}(1313,851)(1333,851)(1352,851)(1371,851)(1390,851)(1410,851)(1429,851)(1448,851)(1467,851)(1487,851)(1506,851)(1525,851)(1544,851)(1564,851)(1583,851)(1602,851)(1621,851)(1641,851)(1660,851)(1679,851)(1698,851)(1718,851)(1737,851)(1756,851)(1775,851)(1795,851)(1814,851)(1833,851)(1852,851)(1872,851)(1891,851)(1910,851)(1929,851)(1949,851)(1968,851)(1987,851)(2006,851)(2026,851)(2045,851)(2064,851)(2083,851)(2103,851)(2122,851)(2141,851)(2160,851)(2180,851)(2199,851)(2218,851)(2237,851)(2257,851)(2276,851)
\thicklines \path(370,249)(2276,249)(2276,1718)(370,1718)(370,249)
\end{picture}
\caption{ The lower bound of $S_{\eta^\prime K}$ of the function of $r$
  for each $\theta $ under constraints for $A_{\eta^\prime K} =
  0.07\pm0.07$, $A_{K\pi} = 0.02\pm0.13$ and $S_{K\pi} = 0.31\pm0.26$,
  where $\delta$ and $w$ are free parameters for {\bf Case B-2}. The
  dotted lines show the experimental data of $S_{\eta^\prime K}$. }
\label{LB}
\end{center}
\end{wrapfigure}
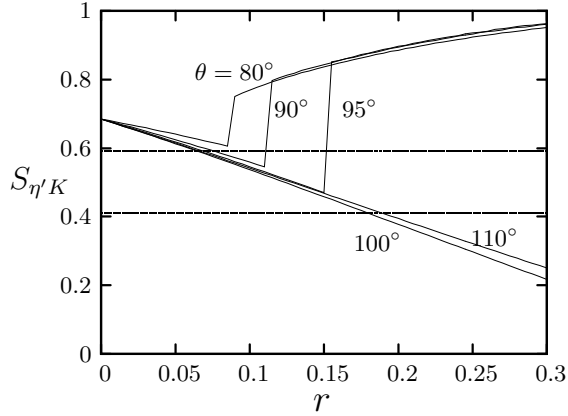
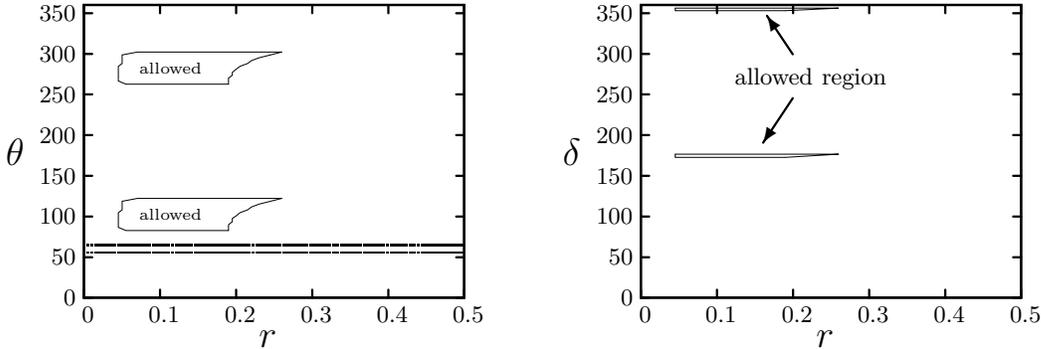
\begin{figure}[htbp]
\begin{center}
\begin{minipage}[l]{2.5in}
\setlength{\unitlength}{0.075450pt}
\begin{picture}(2400,1800)(0,0)
\footnotesize
\thicklines \path(370,249)(411,249)
\thicklines \path(2276,249)(2235,249)
\put(329,249){\makebox(0,0)[r]{ 0}}
\thicklines \path(370,453)(411,453)
\thicklines \path(2276,453)(2235,453)
\put(329,453){\makebox(0,0)[r]{ 50}}
\thicklines \path(370,657)(411,657)
\thicklines \path(2276,657)(2235,657)
\put(329,657){\makebox(0,0)[r]{ 100}}
\thicklines \path(370,861)(411,861)
\thicklines \path(2276,861)(2235,861)
\put(329,861){\makebox(0,0)[r]{ 150}}
\thicklines \path(370,1065)(411,1065)
\thicklines \path(2276,1065)(2235,1065)
\put(329,1065){\makebox(0,0)[r]{ 200}}
\thicklines \path(370,1269)(411,1269)
\thicklines \path(2276,1269)(2235,1269)
\put(329,1269){\makebox(0,0)[r]{ 250}}
\thicklines \path(370,1473)(411,1473)
\thicklines \path(2276,1473)(2235,1473)
\put(329,1473){\makebox(0,0)[r]{ 300}}
\thicklines \path(370,1677)(411,1677)
\thicklines \path(2276,1677)(2235,1677)
\put(329,1677){\makebox(0,0)[r]{ 350}}
\thicklines \path(370,249)(370,290)
\thicklines \path(370,1718)(370,1677)
\put(370,166){\makebox(0,0){ 0}}
\thicklines \path(751,249)(751,290)
\thicklines \path(751,1718)(751,1677)
\put(751,166){\makebox(0,0){ 0.1}}
\thicklines \path(1132,249)(1132,290)
\thicklines \path(1132,1718)(1132,1677)
\put(1132,166){\makebox(0,0){ 0.2}}
\thicklines \path(1514,249)(1514,290)
\thicklines \path(1514,1718)(1514,1677)
\put(1514,166){\makebox(0,0){ 0.3}}
\thicklines \path(1895,249)(1895,290)
\thicklines \path(1895,1718)(1895,1677)
\put(1895,166){\makebox(0,0){ 0.4}}
\thicklines \path(2276,249)(2276,290)
\thicklines \path(2276,1718)(2276,1677)
\put(2276,166){\makebox(0,0){ 0.5}}
\thicklines \path(370,249)(2276,249)(2276,1718)(370,1718)(370,249)
\put(-22,983){\makebox(0,0)[l]{\Large $\theta $}}
\put(1323,42){\makebox(0,0){\Large $r$ }}
\put(821,667){\makebox(0,0){\tiny allowed  }}
\put(821,1400){\makebox(0,0){\tiny allowed  }}
\thinlines \path(1094,587)(1094,587)(1094,602)(1094,616)(1113,631)(1113,646)(1132,660)(1151,675)(1190,690)(1209,704)(1247,719)(1304,734)(1361,748)
\thinlines \path(1094,1321)(1094,1321)(1094,1336)(1094,1351)(1113,1365)(1113,1380)(1132,1395)(1151,1410)(1190,1424)(1209,1439)(1247,1454)(1304,1468)(1361,1483)
\thinlines \path(1094,587)(580,587)(580,587)(542,602)(542,616)(542,631)(542,646)(542,660)(542,675)(561,690)(561,704)(561,719)(561,734)(637,748)(1361,748)
\thinlines \path(1094,1321)(580,1321)(580,1321)(542,1336)(542,1351)(542,1365)(542,1380)(542,1395)(542,1410)(561,1424)(561,1439)(561,1454)(561,1468)(637,1483)(1361,1483)
\thinlines \dashline[-10]{12}(370,476)(370,476)(389,476)(409,476)(428,476)(447,476)(466,476)(486,476)(505,476)(524,476)(543,476)(563,476)(582,476)(601,476)(620,476)(640,476)(659,476)(678,476)(697,476)(717,476)(736,476)(755,476)(774,476)(794,476)(813,476)(832,476)(851,476)(871,476)(890,476)(909,476)(928,476)(948,476)(967,476)(986,476)(1005,476)(1025,476)(1044,476)(1063,476)(1082,476)(1102,476)(1121,476)(1140,476)(1159,476)(1179,476)(1198,476)(1217,476)(1236,476)(1256,476)(1275,476)(1294,476)(1313,476)
\thinlines \dashline[-10]{12}(1313,476)(1333,476)(1352,476)(1371,476)(1390,476)(1410,476)(1429,476)(1448,476)(1467,476)(1487,476)(1506,476)(1525,476)(1544,476)(1564,476)(1583,476)(1602,476)(1621,476)(1641,476)(1660,476)(1679,476)(1698,476)(1718,476)(1737,476)(1756,476)(1775,476)(1795,476)(1814,476)(1833,476)(1852,476)(1872,476)(1891,476)(1910,476)(1929,476)(1949,476)(1968,476)(1987,476)(2006,476)(2026,476)(2045,476)(2064,476)(2083,476)(2103,476)(2122,476)(2141,476)(2160,476)(2180,476)(2199,476)(2218,476)(2237,476)(2257,476)(2276,476)
\thinlines \dashline[-10]{12}(370,513)(370,513)(389,513)(409,513)(428,513)(447,513)(466,513)(486,513)(505,513)(524,513)(543,513)(563,513)(582,513)(601,513)(620,513)(640,513)(659,513)(678,513)(697,513)(717,513)(736,513)(755,513)(774,513)(794,513)(813,513)(832,513)(851,513)(871,513)(890,513)(909,513)(928,513)(948,513)(967,513)(986,513)(1005,513)(1025,513)(1044,513)(1063,513)(1082,513)(1102,513)(1121,513)(1140,513)(1159,513)(1179,513)(1198,513)(1217,513)(1236,513)(1256,513)(1275,513)(1294,513)(1313,513)
\thinlines \dashline[-10]{12}(1313,513)(1333,513)(1352,513)(1371,513)(1390,513)(1410,513)(1429,513)(1448,513)(1467,513)(1487,513)(1506,513)(1525,513)(1544,513)(1564,513)(1583,513)(1602,513)(1621,513)(1641,513)(1660,513)(1679,513)(1698,513)(1718,513)(1737,513)(1756,513)(1775,513)(1795,513)(1814,513)(1833,513)(1852,513)(1872,513)(1891,513)(1910,513)(1929,513)(1949,513)(1968,513)(1987,513)(2006,513)(2026,513)(2045,513)(2064,513)(2083,513)(2103,513)(2122,513)(2141,513)(2160,513)(2180,513)(2199,513)(2218,513)(2237,513)(2257,513)(2276,513)
\thicklines \path(370,249)(2276,249)(2276,1718)(370,1718)(370,249)
\end{picture}
\end{minipage}
    \hspace*{8mm}
\begin{minipage}[r]{2.5in}
\setlength{\unitlength}{0.075450pt}
\begin{picture}(2400,1800)(0,0)
\footnotesize
\thicklines \path(370,249)(411,249)
\thicklines \path(2276,249)(2235,249)
\put(329,249){\makebox(0,0)[r]{ 0}}
\thicklines \path(370,453)(411,453)
\thicklines \path(2276,453)(2235,453)
\put(329,453){\makebox(0,0)[r]{ 50}}
\thicklines \path(370,657)(411,657)
\thicklines \path(2276,657)(2235,657)
\put(329,657){\makebox(0,0)[r]{ 100}}
\thicklines \path(370,861)(411,861)
\thicklines \path(2276,861)(2235,861)
\put(329,861){\makebox(0,0)[r]{ 150}}
\thicklines \path(370,1065)(411,1065)
\thicklines \path(2276,1065)(2235,1065)
\put(329,1065){\makebox(0,0)[r]{ 200}}
\thicklines \path(370,1269)(411,1269)
\thicklines \path(2276,1269)(2235,1269)
\put(329,1269){\makebox(0,0)[r]{ 250}}
\thicklines \path(370,1473)(411,1473)
\thicklines \path(2276,1473)(2235,1473)
\put(329,1473){\makebox(0,0)[r]{ 300}}
\thicklines \path(370,1677)(411,1677)
\thicklines \path(2276,1677)(2235,1677)
\put(329,1677){\makebox(0,0)[r]{ 350}}
\thicklines \path(370,249)(370,290)
\thicklines \path(370,1718)(370,1677)
\put(370,166){\makebox(0,0){ 0}}
\thicklines \path(751,249)(751,290)
\thicklines \path(751,1718)(751,1677)
\put(751,166){\makebox(0,0){ 0.1}}
\thicklines \path(1132,249)(1132,290)
\thicklines \path(1132,1718)(1132,1677)
\put(1132,166){\makebox(0,0){ 0.2}}
\thicklines \path(1514,249)(1514,290)
\thicklines \path(1514,1718)(1514,1677)
\put(1514,166){\makebox(0,0){ 0.3}}
\thicklines \path(1895,249)(1895,290)
\thicklines \path(1895,1718)(1895,1677)
\put(1895,166){\makebox(0,0){ 0.4}}
\thicklines \path(2276,249)(2276,290)
\thicklines \path(2276,1718)(2276,1677)
\put(2276,166){\makebox(0,0){ 0.5}}
\thicklines \path(370,249)(2276,249)(2276,1718)(370,1718)(370,249)
\put(-22,983){\makebox(0,0)[l]{\Large $\delta $}}
\put(1323,42){\makebox(0,0){\Large $r$  }}
\put(1221,1350){\makebox(0,0){ allowed region }}
\put(1130,1470){\vector(-2,3){130}}
\put(1130,1250){\vector(-2,-3){150}}
\thinlines \path(2276,249)(2276,249)
\thinlines \path(1094,954)(1094,954)(1361,969)(542,969)
\thinlines \path(1094,1689)(1094,1689)(1361,1703)(542,1703)
\thinlines \path(542,249)(542,249)
\thinlines \path(1094,954)(542,954)(542,954)(542,969)
\thinlines \path(1094,1689)(542,1689)(542,1689)(542,1703)
\thicklines \path(370,249)(2276,249)(2276,1718)(370,1718)(370,249)
\end{picture}
\end{minipage}
\caption{In {\bf Case B-2}, the allowed region for $r$ and $\theta$
(Left) and $\delta $ (Right) to satisfy $A_{\eta^\prime K},
  S_{\eta^\prime K}, A_{K\pi }$ and $S_{K\pi}$.  Here the other
  parameters are assumed as free. The dotted lines show the
  bound of $\phi_3$ by CKM fitting. }
    \label{fig:new5}
\end{center}
\end{figure}

Using the constraints from Table~\ref{CPdata} ($A_{\eta^\prime K} =
0.07\pm0.07$, $A_{K\pi} = 0.02\pm0.13$ and $S_{K\pi} = 0.31\pm0.26$),
we estimate the $\theta$ and $r$ dependencies by drawing
$S_{\eta^\prime K}$ over $r$ as in Fig.~\ref{LB}. This shows how much
the negative contribution in $S_{\eta^\prime K}$ can be large by the
new weak phase.
Considering the
current experimental bound on $S_{\eta^\prime K}$, we find that
$\theta$ should be larger than about $ 80^\circ$.
In Fig.~\ref{fig:new5}, we also show  the allowed parameter spaces by
using the current experimental data. Here $w$ is treated as a free
parameter. These figures tell us that the allowed region seems to be
quite narrow and $\theta $ should be larger than
the current data of $\phi_3$.
This may indicate that there should be a new weak phase and
agrees with the discussion in {\bf Case B-1}.
We note that $r$ (and $w$) must be also quite larger than the SM
estimates which are ${\cal O}(0.01)$.

Once again, this possibility of having large color-suppressed tree
contribution can help explain several discrepancies, but it may not be
readily acceptable in that large NP effects should be included in the
tree diagram, because NP contributions are usually expected to appear
through some loop effects.

{\bf Case C (with $\bf S_{J/\psi K} \simeq  S_{\eta^\prime K^0}
\neq S_{\phi K} \simeq S_{K^0 \pi^0 } $)-- Possibly Accidental?}

In this case, only the $S_{\eta^\prime K}$ should be different from
the others.  However, since the $B\rightarrow \eta^\prime K$ mode has
all the types of penguin contributions and even color-suppressed tree,
it can not be free from any NP effect appearing in other modes.
Hence, it seems unnatural and complicated to realize this case.  But,
this does not mean that we can simply exclude this case, because it
{\em really} could be controlled by some complicated mechanism.  If we
consider this case, then all $S_f$'s will be different from each
other. In the present work, we do not consider this case any further.
\\

{\bf Case D (with $\bf S_{J/\psi K} \neq  S_{K^0 \pi^0}
\simeq S_{\phi K} \simeq S_{\eta^\prime K} $)}

When we consider the present situation, $S_{\phi K} \simeq
S_{\eta^\prime K} \simeq S_{K^0 \pi^0}$ and $A_{\phi K} \simeq
A_{\eta^\prime K} \simeq A_{K^0 \pi^0} \sim 0 $, it appears that {\bf
  Case D} is the most plausible scenario.  Therefore, we will consider
this case thoroughly.  Neglecting the color-suppressed tree
contribution, the three modes can be parametrized as follows:
\bea
A(B\rightarrow \phi K ) &=& \tilde{A}_{SM} (1 + \tilde{r}
                                                          e^{i\tilde{\delta }}
                                                          e^{i\theta} ), \\
A(B\rightarrow \eta^\prime K^0 ) &=& {A}_{SM} (1 + {r}
                                               e^{i{\delta }}
                                                          e^{i\theta}
                                                          ), \\
A(B\rightarrow \pi^0 K^0 ) &=& {A}^w_{SM} (1 +{w}
                                               e^{i{\delta^w }}
                                               e^{i\theta}
                                                          ),
\eea
where $\tilde{r}, r$ and $w$ are the relative ratios of the NP effects
to the ordinary penguin contributions. Here we assume that the new CP
phase, denoted by $\theta $, is the same for all the modes.
$\tilde{\delta }, \delta $ and $\delta^w $ are the strong phase
differences for each modes.  Using this parametrization, the
time-dependent CP asymmetries are obtained by
\bea
A_{\phi K} &=& \frac{2 \tilde{r} \sin\tilde{\delta} \sin\theta }
                      {1 + \tilde{r}^2 + 2 \tilde{r}
                      \cos\tilde{\delta}
                        \cos\theta }, \\
S_{\phi K} &=& \frac{\sin{2 \phi_1 } + \tilde{r}^2 \sin2(\phi_1+\theta)
                      + 2 \tilde{r} \cos\tilde{\delta}
                          \sin(2 \phi_1 + \theta ) }
                    { 1 + \tilde{r}^2 + 2 \tilde{r} \cos\tilde{\delta
                      } \cos\theta },
                      \\[1mm]
A_{\eta^\prime K} &=& \frac{2 {r} \sin{\delta} \sin\theta }
                      {1 + {r}^2 + 2 {r}
                      \cos{\delta}
                      \cos\theta }, \\
S_{\eta^\prime K} &=& \frac{\sin{2 \phi_1 } + r^2 \sin2(\phi_1+\theta)
                      + 2 r \cos\delta \sin(2 \phi_1 + \theta ) }
                    { 1 + r^2 + 2 r \cos\delta \cos\theta },
                      \\[1mm]
A_{\pi^0 K} &=& \frac{2 w \sin\delta^w \sin\theta }
                      {1 + w^2 + 2 w \cos\delta^w \cos\theta }, \\
S_{\pi^0 K} &=& \frac{\sin{2 \phi_1 } + w^2 \sin2(\phi_1+\theta)
                      + 2 w \cos\delta^w \sin(2 \phi_1 + \theta ) }
                    { 1 + w^2 + 2 w \cos\delta^w \cos\theta }.
\eea
In order to get a practical estimate, let us make some assumptions:
First, considering SU(3) flavor symmetry, we get $P^s = P $.
Moreover, we assume that $ C/T \sim S/P \sim 0.1$ so that the singlet
contribution $S$ is also small enough to be neglected.  Under these
assumptions, we can get some relations for the parameters of
$B\rightarrow \eta^\prime K $ and $B\rightarrow K \pi $. But they may
not relate to those of $B\rightarrow \phi K $ because the one of the
final state is vector meson.

Since the present experimental data are too scarce to determine
precisely all the possible NP effects hidden in $b \to s$ penguins,
here we consider simple assumptions for relations
among the parameters to reduce the number of parameter.
As noticed earlier,  they are only convenient to find roughly the
parameter dependence to
the allowed region for each case.
First, if the NP contribution, say $X$, appears in the QCD penguin
sector\footnote{
Please note that the contribution of $P$ is already including NP contribution of $X$, and
the SM only contribution in $P$ should be $P - X$.
},
the parameters $r$ and $w$ can be estimated as
\bea
      {r} &=& \left|
            \frac{ 3 X }{3 P - 3 X + 4 S - \frac{1}{3} P_{EW} }
                    \right|,   \nn
      \\
       {w} &=& \left| \frac{ X }{P - X - P_{EW} }
                     \right|.   \nn
\eea
Then, they are roughly of the same order of magnitude,
\bea
  r : w = 1 : 1 ,
\eea
so that we can reduce the number of unknown parameters and extract the
information on the four parameters $r=w, \delta,\delta^w$ and $\theta$
from four measurements.  The difference between $B\rightarrow
\eta^\prime K $ and $B \rightarrow K\pi $ comes only from the strong
phases.  If they have the same strong phase, there will be no
difference in the measurements.

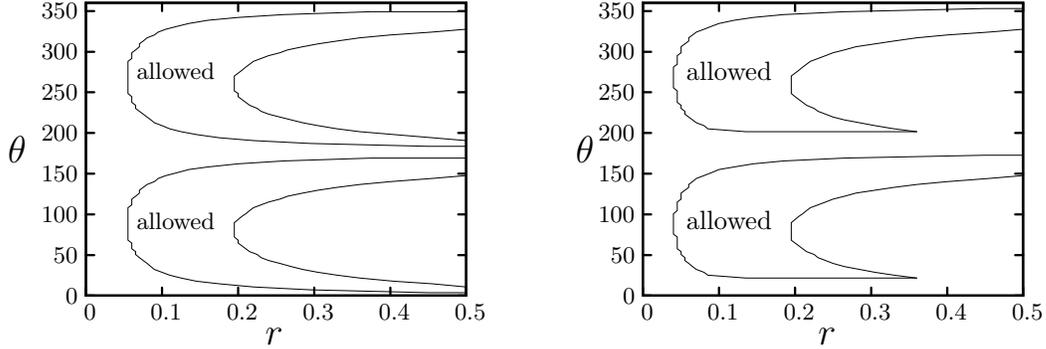
\begin{figure}[t]
\begin{center}
\begin{minipage}[l]{2.5in}
\setlength{\unitlength}{0.075450pt}
\begin{picture}(2400,1800)(0,0)
\footnotesize
\thicklines \path(370,249)(411,249)
\thicklines \path(2276,249)(2235,249)
\put(329,249){\makebox(0,0)[r]{ 0}}
\thicklines \path(370,453)(411,453)
\thicklines \path(2276,453)(2235,453)
\put(329,453){\makebox(0,0)[r]{ 50}}
\thicklines \path(370,657)(411,657)
\thicklines \path(2276,657)(2235,657)
\put(329,657){\makebox(0,0)[r]{ 100}}
\thicklines \path(370,861)(411,861)
\thicklines \path(2276,861)(2235,861)
\put(329,861){\makebox(0,0)[r]{ 150}}
\thicklines \path(370,1065)(411,1065)
\thicklines \path(2276,1065)(2235,1065)
\put(329,1065){\makebox(0,0)[r]{ 200}}
\thicklines \path(370,1269)(411,1269)
\thicklines \path(2276,1269)(2235,1269)
\put(329,1269){\makebox(0,0)[r]{ 250}}
\thicklines \path(370,1473)(411,1473)
\thicklines \path(2276,1473)(2235,1473)
\put(329,1473){\makebox(0,0)[r]{ 300}}
\thicklines \path(370,1677)(411,1677)
\thicklines \path(2276,1677)(2235,1677)
\put(329,1677){\makebox(0,0)[r]{ 350}}
\thicklines \path(370,249)(370,290)
\thicklines \path(370,1718)(370,1677)
\put(370,166){\makebox(0,0){ 0}}
\thicklines \path(751,249)(751,290)
\thicklines \path(751,1718)(751,1677)
\put(751,166){\makebox(0,0){ 0.1}}
\thicklines \path(1132,249)(1132,290)
\thicklines \path(1132,1718)(1132,1677)
\put(1132,166){\makebox(0,0){ 0.2}}
\thicklines \path(1514,249)(1514,290)
\thicklines \path(1514,1718)(1514,1677)
\put(1514,166){\makebox(0,0){ 0.3}}
\thicklines \path(1895,249)(1895,290)
\thicklines \path(1895,1718)(1895,1677)
\put(1895,166){\makebox(0,0){ 0.4}}
\thicklines \path(2276,249)(2276,290)
\thicklines \path(2276,1718)(2276,1677)
\put(2276,166){\makebox(0,0){ 0.5}}
\thicklines \path(370,249)(2276,249)(2276,1718)(370,1718)(370,249)
\put(-22,983){\makebox(0,0)[l]{{\Large $\theta $}}}
\put(1323,42){\makebox(0,0){ \Large $r$  }}
\put(821,627){\makebox(0,0){ allowed  }}
\put(821,1380){\makebox(0,0){ allowed  }}
\thinlines \path(2276,264)(2085,264)(2085,264)(1495,278)(1209,293)(1037,308)(923,322)(847,337)(789,352)(751,367)(713,381)(694,396)(675,411)(656,425)(637,440)(618,455)(618,469)(599,484)(599,499)(599,513)(580,528)(580,543)(580,557)(580,572)(580,587)(580,602)(580,616)(580,631)(580,646)(580,660)(580,675)(580,690)(599,704)(599,719)(599,734)(618,748)(637,763)(637,778)(656,793)(675,807)(713,822)(732,837)(770,851)(827,866)(885,881)(980,895)(1132,910)(1361,925)(1800,939)(2276,939)
\thinlines \path(2276,998)(2085,998)(2085,998)(1495,1013)(1209,1028)(1037,1042)(923,1057)(847,1072)(789,1086)(751,1101)(713,1116)(694,1130)(675,1145)(656,1160)(637,1174)(618,1189)(618,1204)(599,1219)(599,1233)(599,1248)(580,1263)(580,1277)(580,1292)(580,1307)(580,1321)(580,1336)(580,1351)(580,1365)(580,1380)(580,1395)(580,1410)(580,1424)(599,1439)(599,1454)(599,1468)(618,1483)(637,1498)(637,1512)(656,1527)(675,1542)(713,1556)(732,1571)(770,1586)(827,1600)(885,1615)(980,1630)(1132,1645)(1361,1659)(1800,1674)(2276,1674)
\thinlines \path(2276,264)(2276,264)(2276,278)(2276,293)(2104,308)(1914,322)(1742,337)(1628,352)(1533,367)(1456,381)(1399,396)(1342,411)(1285,425)(1247,440)(1228,455)(1190,469)(1171,484)(1151,499)(1132,513)(1132,528)(1113,543)(1113,557)(1113,572)(1113,587)(1113,602)(1113,616)(1132,631)(1151,646)(1171,660)(1190,675)(1209,690)(1247,704)(1285,719)(1342,734)(1380,748)(1456,763)(1533,778)(1628,793)(1742,807)(1895,822)(2104,837)(2276,851)(2276,866)(2276,881)(2276,895)(2276,910)(2276,925)(2276,939)
\thinlines \path(2276,998)(2276,998)(2276,1013)(2276,1028)(2104,1042)(1914,1057)(1742,1072)(1628,1086)(1533,1101)(1456,1116)(1399,1130)(1342,1145)(1285,1160)(1247,1174)(1228,1189)(1190,1204)(1171,1219)(1151,1233)(1132,1248)(1132,1263)(1113,1277)(1113,1292)(1113,1307)(1113,1321)(1113,1336)(1113,1351)(1132,1365)(1151,1380)(1171,1395)(1190,1410)(1209,1424)(1247,1439)(1285,1454)(1342,1468)(1380,1483)(1456,1498)(1533,1512)(1628,1527)(1742,1542)(1895,1556)(2104,1571)(2276,1586)(2276,1600)(2276,1615)(2276,1630)(2276,1645)(2276,1659)(2276,1674)
\thicklines \path(370,249)(2276,249)(2276,1718)(370,1718)(370,249)
\end{picture}
\end{minipage}
    \hspace*{8mm}
\begin{minipage}[r]{2.5in}
\setlength{\unitlength}{0.075450pt}
\begin{picture}(2400,1800)(0,0)
\footnotesize
\thicklines \path(370,249)(411,249)
\thicklines \path(2276,249)(2235,249)
\put(329,249){\makebox(0,0)[r]{ 0}}
\thicklines \path(370,453)(411,453)
\thicklines \path(2276,453)(2235,453)
\put(329,453){\makebox(0,0)[r]{ 50}}
\thicklines \path(370,657)(411,657)
\thicklines \path(2276,657)(2235,657)
\put(329,657){\makebox(0,0)[r]{ 100}}
\thicklines \path(370,861)(411,861)
\thicklines \path(2276,861)(2235,861)
\put(329,861){\makebox(0,0)[r]{ 150}}
\thicklines \path(370,1065)(411,1065)
\thicklines \path(2276,1065)(2235,1065)
\put(329,1065){\makebox(0,0)[r]{ 200}}
\thicklines \path(370,1269)(411,1269)
\thicklines \path(2276,1269)(2235,1269)
\put(329,1269){\makebox(0,0)[r]{ 250}}
\thicklines \path(370,1473)(411,1473)
\thicklines \path(2276,1473)(2235,1473)
\put(329,1473){\makebox(0,0)[r]{ 300}}
\thicklines \path(370,1677)(411,1677)
\thicklines \path(2276,1677)(2235,1677)
\put(329,1677){\makebox(0,0)[r]{ 350}}
\thicklines \path(370,249)(370,290)
\thicklines \path(370,1718)(370,1677)
\put(370,166){\makebox(0,0){ 0}}
\thicklines \path(751,249)(751,290)
\thicklines \path(751,1718)(751,1677)
\put(751,166){\makebox(0,0){ 0.1}}
\thicklines \path(1132,249)(1132,290)
\thicklines \path(1132,1718)(1132,1677)
\put(1132,166){\makebox(0,0){ 0.2}}
\thicklines \path(1514,249)(1514,290)
\thicklines \path(1514,1718)(1514,1677)
\put(1514,166){\makebox(0,0){ 0.3}}
\thicklines \path(1895,249)(1895,290)
\thicklines \path(1895,1718)(1895,1677)
\put(1895,166){\makebox(0,0){ 0.4}}
\thicklines \path(2276,249)(2276,290)
\thicklines \path(2276,1718)(2276,1677)
\put(2276,166){\makebox(0,0){ 0.5}}
\thicklines \path(370,249)(2276,249)(2276,1718)(370,1718)(370,249)
\put(32,983){\makebox(0,0)[l]{\Large{$\theta $}}}
\put(1323,42){\makebox(0,0){\Large $ r$  }}
\put(821,627){\makebox(0,0){\small allowed  }}
\put(821,1380){\makebox(0,0){\small allowed  }}
\thinlines \path(1742,337)(885,337)(885,337)(694,352)(675,367)(637,381)(618,396)(599,411)(580,425)(580,440)(561,455)(561,469)(542,484)(542,499)(542,513)(542,528)(542,543)(522,557)(522,572)(522,587)(522,602)(522,616)(522,631)(522,646)(522,660)(542,675)(542,690)(542,704)(542,719)(561,734)(561,748)(561,763)(580,778)(599,793)(599,807)(618,822)(637,837)(675,851)(713,866)(751,881)(827,895)(923,910)(1075,925)(1380,939)(2085,954)(2276,954)
\thinlines \path(1742,1072)(885,1072)(885,1072)(694,1086)(675,1101)(637,1116)(618,1130)(599,1145)(580,1160)(580,1174)(561,1189)(561,1204)(542,1219)(542,1233)(542,1248)(542,1263)(542,1277)(522,1292)(522,1307)(522,1321)(522,1336)(522,1351)(522,1365)(522,1380)(522,1395)(542,1410)(542,1424)(542,1439)(542,1454)(561,1468)(561,1483)(561,1498)(580,1512)(599,1527)(599,1542)(618,1556)(637,1571)(675,1586)(713,1600)(751,1615)(827,1630)(923,1645)(1075,1659)(1380,1674)(2085,1689)(2276,1689)
\thinlines \path(1742,337)(1742,337)(1628,352)(1533,367)(1456,381)(1380,396)(1323,411)(1285,425)(1247,440)(1228,455)(1190,469)(1171,484)(1151,499)(1132,513)(1113,528)(1113,543)(1113,557)(1113,572)(1113,587)(1113,602)(1113,616)(1132,631)(1151,646)(1171,660)(1190,675)(1209,690)(1247,704)(1285,719)(1323,734)(1380,748)(1437,763)(1533,778)(1628,793)(1742,807)(1895,822)(2104,837)(2276,851)(2276,866)(2276,881)(2276,895)(2276,910)(2276,925)(2276,939)(2276,954)
\thinlines \path(1742,1072)(1742,1072)(1628,1086)(1533,1101)(1456,1116)(1380,1130)(1323,1145)(1285,1160)(1247,1174)(1228,1189)(1190,1204)(1171,1219)(1151,1233)(1132,1248)(1113,1263)(1113,1277)(1113,1292)(1113,1307)(1113,1321)(1113,1336)(1113,1351)(1132,1365)(1151,1380)(1171,1395)(1190,1410)(1209,1424)(1247,1439)(1285,1454)(1323,1468)(1380,1483)(1437,1498)(1533,1512)(1628,1527)(1742,1542)(1895,1556)(2104,1571)(2276,1586)(2276,1600)(2276,1615)(2276,1630)(2276,1645)(2276,1659)(2276,1674)(2276,1689)
\thicklines \path(370,249)(2276,249)(2276,1718)(370,1718)(370,249)
\end{picture}
\end{minipage}
\caption{For {\bf Case D}, the allowed region for $r$ and $\theta $ to satisfy
$A_{\eta^\prime K}, S_{\eta^\prime K}, A_{K\pi }$ and $S_{K\pi }$
under relations $ r = w $ (Left), which shows it comes from
penguin diagrams, and $ 9 r \sim w $ (Right), which comes from EW penguin.  }
\label{fig:6}
\end{center}
\end{figure}

If NP shows up in the EW penguin sector, then
\bea
      {r} &=& \left| \frac{ \frac{1}{3}X }{3 P + 4 S - \frac{1}{3} (P_{EW}
      -X ) } \right|, \nn
      \\
       {w} &=& \left| \frac{ X }{P - ( P_{EW}- X ) }\right|.  \nn
\eea
Roughly speaking, in this case,
\bea
    r : w &=& 1 : \left|\frac{9P + 12S - (P_{EW} - X )}
                           {P-(P_{EW}- X)} \right| \nn \\
     &\sim & 1 : \frac{9}{\left|1 + \frac{8X}{9P+X} \right|} =
1 : \frac{9}{\sqrt{1+64r^2 + 16 r \cos\delta \cos\theta }}
\eea
where $r$ is almost $|X/(9P)|$.  For the above two cases with $
\frac{w}{r} = 1 $ and $\frac{w}{r} =\frac{9}{\sqrt{1+64r^2 + 16 r
    \cos\delta \cos\theta }}$, the allowed regions for $r$ and
$\theta$ are plotted in Fig.~\ref{fig:6}.

Our estimate shows that NP contributions must have a large weak
phase to explain the present discrepancies. However, we need more precise
experimental results to single out, hence understand the CP
violation effects due to NP.  Though our present analysis seems to be quite
rough, at least, we expect one may find which case will
be preferred by the results from $B$-factory experiments in near
future. To extract the more definite hint about NP effects,
we may have to add some extra information about the parameters
from the other experimental data. In next section, we discuss
how to classify the new physics contributions
by adding the information of $B_s$ decays.

\section{Complementary Analysis through $B_s$ and $B_d$ decays}

If the discrepancy in the indirect CP asymmetry $S_f$ is true, it will
be more important to know where the source of the discrepancy is.  As
shown in the previous section, the differences among the $S_f$'s of
$B\rightarrow \phi K$, $B\rightarrow \eta^\prime K$ and $B\rightarrow
K\pi $ could lead to some information about NP effects. However, if
the situation is {\bf Case D}, then we should be able to separate each
contribution to find where the NP effects may come from.  In this
regard, using the time-dependent CP asymmetries of the $B_s$ decays,
one can obtain further useful information.  As listed in
Table~\ref{Bd-Bs}, the $B_s \rightarrow K^0\overline{K}^0 $ mode is
almost a pure QCD penguin process, and the $B_s \rightarrow
\eta^\prime {\pi}^0 $ is almost a pure EW penguin mode. If one of them
is including a large weak phase, its effect should appear in the
$S_f$.  Therefore, any sizable difference between
$S_{K^0\overline{K}^0} $ and $S_{\eta^\prime K^0}$ will directly imply
the different origin of the weak phases.  The two processes can be
expressed as
%
\newcommand{\bkk}{ \langle K^0 \overline{K}^0 \rangle }
\newcommand{\bep}{ \langle \eta^\prime {\pi }^0 \rangle  }
\bea
A(B_s \rightarrow K^0 \overline{K}^0) &=& P V_{tb}^* V_{ts}
                        \equiv |P V_{tb}^* V_{ts}| e^{i\delta }
                              e^{i{\phi_{D {\bkk}}}},  \\
A(B_s \rightarrow \eta^\prime {\pi}^0) &=& \frac{2}{\sqrt{6}}
                                   P_{EW} V_{tb}^* V_{ts}
                   \equiv \frac{2}{\sqrt{6}}|P_{EW} V_{tb}^* V_{ts}|
                              e^{i\delta }
                              e^{i{\phi_{D \bep}}}.
\eea
Then, the indirect CP asymmetries are given as $S_{K^0 \overline{K}^0}^s
        = \sin{2{\phi_{D \bkk}^s} }$ and $S_{\eta^\prime \pi^0}^s
        = \sin{2{\phi_{D \bep}^s} }$.
The difference between $S_{K^0\overline{K}^0}^s $ and
$S_{\eta^\prime K^0}^s$ will lead directly to the difference
between the angles ${\phi_{D \bkk} }$ and ${\phi_{D \bep}}$,
which may have different origins.  To be consistent with the
situation of Fig.~\ref{F1}, one or both of them should have
non-negligible value under SU(3) flavor symmetry.  Even if
some new phase also comes in the $B_s-\overline{B}_s$ mixing
process, nevertheless the difference will appear between them.
In Fig. 6, we summarize possible roles of the $B_s$ decays to
classify the types of NP.
\begin{figure}[t]
\label{fig:flowchart}
\begin{center}
\unitlength 3.8mm
{\footnotesize
\begin{picture}(30,11)
\put(9.5,10){\fbox{{\bf Case D}: ~~ $S_{\phi K} \sim S_{\eta^\prime K } \sim
        S_{K\pi } $}}
\put(15,9.5){\vector(0,-3){1}}
\put(15,8.4){\line(-3,0){8}}
\put(15,8.4){\line(3,0){8}}
\put(7,8.4){\vector(0,-2){1}}
\put(23,8.4){\vector(0,-2){1}}
\put(4,6.5){\fbox{ $S_{K^0\overline{K}^0}^s \neq S_{\eta^\prime \pi^0 }^s $}}
\put(20,6.5){\fbox{ $ S_{K^0\overline{K}^0}^s = S_{\eta^\prime \pi^0 }^s$}}
\put(7,5.9){\vector(0,-3){1}}
\put(23,5.9){\vector(0,-3){1}}
\put(7,5){\line(-3,0){5}}
\put(7,5){\line(3,0){5}}
\put(23,5){\line(-3,0){5}}
\put(23,5){\line(3,0){5}}
\put(2,5){\vector(0,-2){1}}
\put(12,5){\vector(0,-2){1}}
\put(18,5){\vector(0,-2){1}}
\put(28,5){\vector(0,-2){1}}
\put(-2.8,3.0){\fbox{ $S_{K^0\overline{K}^0}^s \neq S_{\eta^\prime \pi^0 }^s \sim
        0 $}}
\put(6.6,3.0){\fbox{ $0 \sim S_{K^0\overline{K}^0}^s \neq S_{\eta^\prime
\pi^0 }^s$}}
\put(15.2,3.0){\fbox{ $S_{K^0\overline{K}^0}^s = S_{\eta^\prime \pi^0 }^s \sim
        0 $}}
\put(24.2,3.0){\fbox{ $S_{K^0\overline{K}^0}^s = S_{\eta^\prime \pi^0 }^s
\neq 0$}}
\put(2,2.4){\vector(0,-2){1}}
\put(12,2.4){\vector(0,-2){1}}
\put(18,2.4){\vector(0,-2){1}}
\put(28,2.4){\vector(0,-2){1}}
\put(-2.8,0.5){\fbox{ QCD penguin type }}
\put(6.6,0.5){\fbox{ EW penguin type }}
\put(15.2,0.5){\fbox{No phase in penguin }}
\put(23.9,0.5){\fbox{\parbox{1.5cm}{\tiny QCD and \\EW penguins} or $B_s $ mixing }}
\end{picture}
}
\caption{Flowchart to classify the type of New Physics in {\bf Case D}. }
\end{center}
\end{figure}
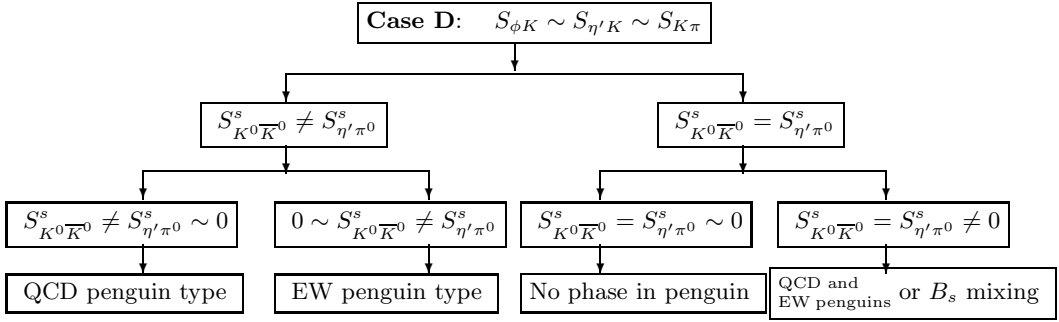

Similar analysis can be applied to the $B_d \rightarrow K^0\overline{K}^0$
mode which is a pure $b\to d$ QCD penguin process.
The decay amplitude is expressed as follows:
\bea
A(B_d \rightarrow K^0 \overline{K}^0) &=& P V_{tb}^* V_{td}
                        \equiv |P V_{tb}^* V_{td}| e^{i\delta }
                              e^{i\left(- \phi_1 + {\phi^d_{D \bkk } }\right)}.
\eea
Then, the time-dependent CP asymmetry $S_{K^0 \overline{K}^0}^d$ is
obtained by
\bea
S_{K^0 \overline{K}^0}^d = \sqrt{ 1 - A^2 }  Im \left[ e^{-2i\phi_1}
           \frac{ A({B}^0 \rightarrow K^0 \overline{K}^0 )^* }
                {|A({B}^0 \rightarrow K^0 \overline{K}^0 )|}
           \frac{ A(\overline{B}^0 \rightarrow K^0 \overline{K}^0 ) }
                {|A(\overline{B}^0 \rightarrow K^0 \overline{K}^0 )|} \right]
           = \sin(2 {\phi_{D \bkk }^d }), \nn \\
\eea
when the weak phase in the $B_d-\overline{B}_d$ mixing is the same as
the SM expectation $\phi_1$.
Therefore, the observation of a sizable
$S_{K^0 \overline{K}^0}^d$ may provide an evidence for the corresponding
NP effect in $b\to d$ QCD penguin sector.
Furthermore, comparing $S_{K^0 \overline{K}^0}^d$
and $S_{K^0 \overline{K}^0}^s$, one can check whether the source of the new CP
phase is in $b\to s$ or $b\to d$ penguin sector, or possibly in both sectors.
Hence, these modes will be very important to understand in which processes
new CP phases might exist.

In the actual analysis, however, there could be more uncertainties to
extract the information than a naive expectation.  To be more
specific, let us consider $B_d \rightarrow K^0 \overline{K}^0$ mode
somewhat in detail.  In the $b\to d$ penguin process, there are three
diagrams with different internal particles in the loop.  Using the
unitarity relation of CKM matrix, the amplitude will be decomposed as
follows:
\bea
A(B_d \rightarrow K^0 \overline{K}^0) &=& (P_t - P_u) V_{tb}^* V_{td}
                                    + (P_c - P_u) V_{cb}^* V_{cd} \nn \\
               &\equiv & |P_{tu} V_{tb}^* V_{td}| e^{i\delta }
                              e^{i(- \phi_1 + {\phi_D}_t )}
                        - |P_{cu} V_{cb}^* V_{cd}| e^{i\delta }
                              e^{i({\phi_D}_c)},
\eea
where we assumed the direct CP asymmetry is absent so that there is no strong
phase difference between the two terms.
Then, we have
\bea
S_{K^0 \overline{K}^0}^d \equiv \sin(2{\phi_{D \bkk}^d})
=\sin2( Z - \phi_1 ),
\eea
where
\bea
\tan Z &=& \frac{ |P_{tu} V_{tb}^* V_{td}| \sin(\phi_1 - {\phi_D}_t)
                + |P_{cu} V_{cb}^* V_{cd}| \sin(- {\phi_D}_u) }
              { |P_{tu} V_{tb}^* V_{td}| \cos(\phi_1 - {\phi_D}_t)
                - |P_{cu} V_{cb}^* V_{cd}| \cos({\phi_D}_u) } \nn
              \\
       &=& \tan(\phi_1 -{\phi_D}_t)
          + \frac{ |P_{cu} V_{cb}^* V_{cd}| }{|P_{tu} V_{tb}^*
              V_{td}|} \frac{\sin(\phi_1-{\phi_D}_t+{\phi_D}_u)}
                            {\cos^2(\phi_1-{\phi_D}_t)}
                          + O\left( (P_{cu}/P_{tu})^2 \right).
\eea Treating $(P_{cu}/P_{tu})$ as a small parameter~\cite{SmCP},
the angle extracted from the time-dependent CP asymmetry will be
\bea {\phi_{D \bkk}^d }\sim {\phi_D}_t . \eea Considering SU(3)
flavor symmetry, one can expect that ${\phi_D}_t$ is also related
to the angle ${\phi_{D \bkk}^s }$ extracted from $B_s\rightarrow
K^0 \overline{K}^0 $. So we can check the consistency by
comparing each other. However, we note that the analysis can be
more difficult if the charming penguin contribution is not small
enough to be simply neglected.

\section{Discussions and Conclusions}

Time-dependent CP asymmetries for several $B^0$ decay modes may
include some fruitful information about weak phases, especially, from
New Physics. Actually, the current experimental results from the $B$
factories seem to be indicating a possibility of the presence of
some NP effects in the $b\to s$ penguin-dominated modes. As $B$ factory
experiments are accumulating data, the measurements are getting more
precise and with probing more and more channels. So, in the near future,
it is expected that the discrepancies will be confirmed or possibly
turned out to be no discrepancy at an acceptable confidence level. If
the {\it currently-appearing} discrepancy is real, as a next step, we
should consider how to determine which type of interactions it may come
from. Although there are still remaining uncertainties, considering
Belle and BaBar experimental results and taking the average over all
the possible measurements with the same SM expectation, some
discrepancies seem to be getting manifest as shown in the recent
averaged data~\cite{HFAG}. But it is not sure whether one can take the
average among all the $b\to s$ penguin type modes, because the
dependence of the diagrams may be somewhat different. In fact, in
order to discuss the NP effects we have to distinguish the related
modes by topological diagrammatic decomposition method.

In this work, we have considered $B \rightarrow \phi K^0$, $B
\rightarrow \eta^\prime K^0$ and $B \rightarrow K^0 \pi^0$ decay
modes. We have derived a systematic classification on possible NP
dependencies appearing in time-dependent CP asymmetries and also
investigated which type of diagrams should be including such new
phases. According to the current experimental results for the three
modes, there is no large discrepancy in the $S_f$'s within those three
modes.  If this situation continues to remain even after the data are
sufficiently updated, it may be natural to infer that the new
contribution should reside in their common diagrams. Under the present
situation, it is difficult to extract more details about the common
contributions because all the three modes are including QCD and EW
penguins. In this regard, we have shown that $B_s$ decay modes can be
very useful. Using the time-dependent CP asymmetries of $B_s$ decay
modes such as $B_s \rightarrow \phi \eta^\prime $, $B_s\rightarrow
\eta^\prime \pi^0$ and $B_s\rightarrow K^0 \overline{K}^0$, one may
determine where the new CP phase comes from.  If there are some
discrepancies among the $B_s$ modes, it can directly imply the
different origin of the new CP phase, even if the $B_s-\overline{B}_s$
mixing includes a new weak phase.

In the future hadron-collision experiments such as ATLAS and CMS
at LHC, the production cross-section of $B_s$ will be enormous,
hence they will provide opportunities to study $B_s$ decays in
great details. For the above-mentioned decays $B_s \rightarrow
\phi \eta^\prime $ and $B_s\rightarrow \eta^\prime \pi^0$,
however, it is essential to detect $\pi^0$'s and photons with a
good precision. Experimental environments of LHC may not be suited
for this purpose. On the other hand, the $e^+ e^-$ $B$-factory
experiments, BaBar and Belle, show very good performances in the
studies of final states involving $\pi^0$ and/or single photons.
To study $B_s$ in the $e^+ e^-$ collision, viable options are
running at $Z^0$ or $\Upsilon(5S)$ resonance energies. But with
LHC in operation, reviving the LEP for running at $Z^0$ resonance
will be practically out of the question. In contrast, it may not
be of much problem to change the beam energies of the current $B$
factories for operating at $\Upsilon(5S)$. Recently, CLEO measured
that ${\cal B}(\Upsilon(5S)\rightarrow B_s^{(*)}
\overline{B}_s^{(*)})=(21 \pm 3 \pm 9)\%$~\cite{CLEO5S}. This,
combined with $\sigma(e^+ e^- \rightarrow \Upsilon(4S))\approx
3\times \sigma(e^+ e^- \rightarrow \Upsilon(5S))$, tells us that
we need approximately 15 times more integrated luminosity
operating at $\Upsilon(5S)$ in order to obtain experimental
sensitivities for $B_s$ decays corresponding to those of similar
$B_d$ decays. The so-called ``Super-$B$ factory'', with more than
an order of magnitude improvement in the instantaneous luminosity
appears to be indispensable for this study.

Moreover, we have also discussed that $B_d \rightarrow K^0
\overline{K}^0$ mode can be used to probe the different dependence
between $b\to s$ and $b\to d$ transition systems. Since the $B_d
\rightarrow K^0\overline{K}^0$ mode is a pure QCD penguin process,
one can investigate the NP effects in the QCD penguin sector
alone. As the results for this mode are expected to appear in $B$
factories, possibly after upgraded to super-$B$ factory
project~\cite{LOI}, we will get more fruitful information on CP
physics in penguin sector in the near future.

\section*{Acknowledgments}

The work of C.S.K. was supported
in part by  CHEP-SRC Program and
in part by the Korea Research Foundation Grant funded by the Korean Government
(MOEHRD) No. R02-2003-000-10050-0.
The work of Y.J.K. was supported
by the Korea Research Foundation Grant funded by the Korean Government
(MOEHRD) No. KRF-2005-070-C00030.
The work of J.L. was supported by
the SRC program of KOSEF through
CQUeST with Grant No. R11-2005-021.
The work of T.Y. was supported by 21st Century COE Program of
Nagoya University provided by JSPS.\\

\end{document}